\documentstyle[12pt,indent,epsf,eqsection,subeqnarray]{article}
\begin{document}

\bibliographystyle{plain}

\date{May 14, 1996 \\ revised February 11, 1997}

\title{Dynamic Critical Behavior of the \break 
       Swendsen--Wang Algorithm: \break  
       The Two-Dimensional 3-State Potts Model Revisited  
      }
\author{
  {\small Jes\'us Salas\thanks{Current address for Salas:
      Departamento de F\'{\i}sica Te\'orica,
      Facultad de Ciencias, Universidad de Zaragoza,
      Zaragoza 50009, SPAIN;  {\tt JESUS@JUPITER.UNIZAR.ES}.}
}                  \\[-0.2cm]
  {\small Alan D.~Sokal}                  \\[-0.2cm]
  {\small\it Department of Physics}       \\[-0.2cm]
  {\small\it New York University}         \\[-0.2cm]
  {\small\it 4 Washington Place}          \\[-0.2cm]
  {\small\it New York, NY 10003 USA}      \\[-0.2cm]
  {\small\tt SALAS@MAFALDA.PHYSICS.NYU.EDU},
                    {\small\tt SOKAL@NYU.EDU}   \\[-0.2cm]
  {\protect\makebox[5in]{\quad}}  % To force authors' names to be written
                                  %   vertically, one above another.
                                  % (\author seems to put them side-by-side
                                  %   if there is room.)
  \\
}
\vspace{0.5cm}

\maketitle
\thispagestyle{empty}   % Suppress page number on front page.

%\ltapprox and \gtapprox produce > and < signs with twiddle underneath
\def\spose#1{\hbox to 0pt{#1\hss}}
\def\ltapprox{\mathrel{\spose{\lower 3pt\hbox{$\mathchar"218$}}
 \raise 2.0pt\hbox{$\mathchar"13C$}}}
\def\gtapprox{\mathrel{\spose{\lower 3pt\hbox{$\mathchar"218$}}
 \raise 2.0pt\hbox{$\mathchar"13E$}}}
\def\inapprox{\mathrel{\spose{\lower 3pt\hbox{$\mathchar"218$}}
 \raise 2.0pt\hbox{$\mathchar"232$}}}

%%\doublespace

\begin{abstract}
We have performed a high-precision Monte Carlo study of the dynamic 
critical behavior of the Swendsen--Wang algorithm for the two-dimensional
3-state Potts model. We find that the Li--Sokal bound 
($\tau_{{\rm int},{\cal E}} \geq {\rm const} \times C_H$) is almost 
but not quite sharp. The ratio $\tau_{{\rm int},{\cal E}}/C_H$ seems 
to diverge either as a small power ($\approx 0.08$) or as a logarithm. 
\end{abstract}


\clearpage

\newcommand{\be}{\begin{equation}}
\newcommand{\ee}{\end{equation}}
\newcommand{\<}{\langle}
\renewcommand{\>}{\rangle}
\newcommand{\para}{\|}
\renewcommand{\perp}{\bot}

\def\smfrac#1#2{{\textstyle\frac{#1}{#2}}}
\def\half{ {{1 \over 2 }}}
\def\smhalf{ {\smfrac{1}{2}} }
\def\scra{{\cal A}}
\def\scrc{{\cal C}}
\def\scrd{{\cal D}}
\def\scre{{\cal E}}
\def\scrf{{\cal F}}
\def\scrh{{\cal H}}
\def\scrk{{\cal K}}
\def\scrm{{\cal M}}
\newcommand{\scrmvec}{\vec{\cal M}_V}
\def\scrmtens{{\stackrel{\leftrightarrow}{\cal M}_T}}
\def\scro{{\cal O}}
\def\scrp{{\cal P}}
\def\scrr{{\cal R}}
\def\scrs{{\cal S}}
\def\ttens{{\stackrel{\leftrightarrow}{T}}}
\def\scrv{{\cal V}}
\def\scrw{{\cal W}}
\def\scry{{\cal Y}}
\def\tauss{\tau_{int,\,\scrm^2}}
\def\taux{\tau_{int,\,{\cal M}^2}}
\newcommand{\taum}{\tau_{int,\,\vec{\cal M}}}
\def\taue{\tau_{int,\,{\cal E}}}
\newcommand{\imag}{\mathop{\rm Im}\nolimits}
\newcommand{\real}{\mathop{\rm Re}\nolimits}
\newcommand{\tr}{\mathop{\rm tr}\nolimits}
\newcommand{\sgn}{\mathop{\rm sgn}\nolimits}
\newcommand{\codim}{\mathop{\rm codim}\nolimits}
\newcommand{\rank}{\mathop{\rm rank}\nolimits}
\newcommand{\sech}{\mathop{\rm sech}\nolimits}
\def\textprime{{${}^\prime$}}
\newcommand{\longto}{\longrightarrow}
\def\var{ \hbox{var} }
\newcommand{\gtilde}{ {\widetilde{G}} }
\newcommand{\USp}{ \hbox{\it USp} }
\newcommand{\CP}{ \hbox{\it CP\/} }
\newcommand{\QP}{ \hbox{\it QP\/} }
\def\hboxscript#1{ {\hbox{\scriptsize\em #1}} }

\newcommand{\plotdot}{\makebox(0,0){$\bullet$}}
\newcommand{\plotsmalldot}{\makebox(0,0){{\footnotesize $\bullet$}}}

\def\bsigma{\mbox{\protect\boldmath $\sigma$}}
\def\bpi{\mbox{\protect\boldmath $\pi$}}
\def\btau{\mbox{\protect\boldmath $\tau$}}
  % \boldmath is fragile, and without the \protect we get screwed when
  % we try to use \bsigma in a \caption.
\def\bn{{\bf n}}
\def\br{{\bf r}}
\def\bz{{\bf z}}
\def\bh{\mbox{\protect\boldmath $h$}}

\def\betatilde{ {\widetilde{\beta}} }
\def\hatp{\hat p}
\def\hatl{\hat l}

\def\msbar{ {\overline{\hbox{\scriptsize MS}}} }
\def\normalmsbar{ {\overline{\hbox{\normalsize MS}}} }

\def\eff{ {\hbox{\scriptsize\em eff}} }

\newcommand{\reff}[1]{(\ref{#1})}

\font\specialroman=msym10 scaled\magstep1  % 12-point Special Roman (caps only)
\font\sevenspecialroman=msym7              % 7-point Special Roman (caps only)
\def\Z{\hbox{\specialroman Z}}
\def\zed{\hbox{\specialroman Z}}
\def\szed{\hbox{\sevenspecialroman Z}}
\def\R{\hbox{\specialroman R}}
\def\sR{\hbox{\sevenspecialroman R}}
\def\N{\hbox{\specialroman N}}
\def\C{\hbox{\specialroman C}}
\def\Q{\hbox{\specialroman Q}}
\renewcommand{\emptyset}{\hbox{\specialroman ?}}
%\newcommand{\Z}{{\bf Z}}
%\newcommand{\zed}{{\bf \Z}}
%\newcommand{\R}{\hbox{{\rm I}\kern-.2em\hbox{\rm R}}}
%\font\srm=cmr7 		% to get seven roman
%\def\szed{\hbox{\srm Z\kern-.45em\hbox{\srm Z}}}
%\def\sR{\hbox{{\srm I}\kern-.2em\hbox{\srm R}}}
%\def\C{{\bf C}}

% \font\german=eufm10 scaled\magstep1	% 12-point Euler Fraktur (German)
% \def\germang{\hbox{\german g}}
% \def\germansu{\hbox{\german su}}

% \font\amssymbol=msxm10 scaled \magstep1  % Another AMS symbol font
% \def\transversal{\hbox{\amssymbol t}}  % THERE MAY BE A BETTER SYMBOL.

\newtheorem{theorem}{Theorem}[section]
\newtheorem{corollary}[theorem]{Corollary}
\newtheorem{lemma}[theorem]{Lemma}
\newtheorem{conjecture}[theorem]{Conjecture}
\newtheorem{definition}[theorem]{Definition}
\def\proof{\bigskip\par\noindent{\sc Proof.\ }}
\def\qed{\hbox{\hskip 6pt\vrule width6pt height7pt depth1pt \hskip1pt}\bigskip}

%
% Array for subscripts
%
\newenvironment{sarray}{
          \textfont0=\scriptfont0
          \scriptfont0=\scriptscriptfont0
          \textfont1=\scriptfont1
          \scriptfont1=\scriptscriptfont1
          \textfont2=\scriptfont2
          \scriptfont2=\scriptscriptfont2
          \textfont3=\scriptfont3
          \scriptfont3=\scriptscriptfont3
        \renewcommand{\arraystretch}{0.7}
        \begin{array}{l}}{\end{array}}

\newenvironment{scarray}{
          \textfont0=\scriptfont0
          \scriptfont0=\scriptscriptfont0
          \textfont1=\scriptfont1
          \scriptfont1=\scriptscriptfont1
          \textfont2=\scriptfont2
          \scriptfont2=\scriptscriptfont2
          \textfont3=\scriptfont3
          \scriptfont3=\scriptscriptfont3
        \renewcommand{\arraystretch}{0.7}
        \begin{array}{c}}{\end{array}}

%%%%%%%%%%%%% BEGINNING OF THE TEXT %%%%%%%%%%%%%%%%%%%%%%%%%%%%%%%%%%%%

\section{Introduction}  \label{sec_intro}

Monte Carlo (MC) simulations  
\cite{Binder_78,Binder_87,Binder_92,Sokal_Lausanne} 
have become a standard and powerful tool for gaining new insights into 
statistical-mechanical systems and lattice field theories. 
However, their practical success is severely limited by critical 
slowing-down: the autocorrelation time $\tau$ --- that is, roughly speaking,  
the time needed to produce one ``statistically independent'' configuration ---  
diverges near a critical point.  
More precisely, for a finite system of linear size $L$ at criticality,
we expect a behavior $\tau \sim L^{z}$ for large $L$.
The power $z$ is a {\em dynamic critical exponent}\/,
and it depends on both the system and the algorithm.

Single-site MC algorithms (such as single-site Metropolis or heat bath)
have a dynamic critical exponent $z \gtapprox 2$. 
This makes it very hard to get high-precision data very close to the 
critical point on large lattices.

In some cases, a much better dynamical behavior is obtained by allowing 
non-local moves, such as cluster flips.\footnote{ 
   See \cite{Sokal_Lausanne,Wolff_LAT89,Sokal_LAT90} for reviews of  
   collective-mode Monte Carlo methods. 
}  
The Swendsen--Wang (SW) cluster algorithm \cite{Swendsen_87} for the 
$q$-state ferromagnetic Potts model achieves 
a significant reduction in $z$ compared to the local algorithms: one has 
$z$ between 0 and $\approx\! 1$, 
where the exact value depends on $q$ and on the dimensionality of the  
lattice \cite{Sokal_LAT90}.  
The most favorable case is the two-dimensional (2D) Ising model ($q=2$), in  
which $z \ltapprox 0.3$ \cite{Swendsen_87,Heermann_90,Baillie_91,Baillie_92} 
(see below). 
In other cases, the performance of the SW algorithm is less impressive 
(though still quite good): e.g., $z = 0.55 \pm 0.03$ for the 2D 3-state 
Potts model \cite{Li_Sokal}, $z \approx 1$ for the 2D 4-state Potts model
\cite{Li_Sokal,Salas_Sokal_LAT95,Salas_Sokal_AT,Salas_Sokal_FSS}, 
and $z \approx 1$  
for the 4D Ising model 
\cite{Klein_89,Ray_89}. 
Clearly, we would like to understand why this algorithm works so well
in some cases and not in others.   
We hope in this way to obtain new insights into the
dynamics of non-local Monte Carlo algorithms,
with the ultimate aim of devising new and more efficient algorithms.

There is at present no adequate theory for predicting the
dynamic critical behavior of an SW-type algorithm.  
However, there is one rigorous lower bound on $z$ due to Li and 
Sokal \cite{Li_Sokal}.  
The autocorrelation times of the standard (multi-cluster)
SW algorithm for the ferromagnetic $q$-state Potts model  
are bounded below by a multiple of the specific heat:
\be 
  \tau_{{\rm int},{\cal N}}, \;
  \tau_{{\rm int},{\cal E}}, \;
  \tau_{\rm exp}  \;\geq\; {\rm const} \times C_H   \;.
\label{Li_Sokal_bound_CH} 
\ee 
Here ${\cal N}$ is the bond density in the SW algorithm,
${\cal E}$ is the energy, and $C_H$ is the specific heat;
$\tau_{\rm int}$ and $\tau_{\rm exp}$ denote the integrated and
exponential autocorrelation times, respectively
\cite{Sokal_Lausanne,Sokal_LAT90}.
As a result one has
\be 
  z_{{\rm int},{\cal N}}, \;
  z_{{\rm int},{\cal E}}, \;
  z_{\rm exp}  \;\geq\; {\alpha \over \nu } \; ,  
\label{Li_Sokal_bound} 
\ee 
where $\alpha$ and $\nu$ are the standard {\em static}\/ critical exponents. 
Thus, the SW algorithm for the $q$-state Potts model cannot 
{\em completely}\/ eliminate the critical slowing-down in any model
in which the specific heat is divergent at criticality. 
The bound \reff{Li_Sokal_bound_CH}/\reff{Li_Sokal_bound} has also been 
proven to hold \cite{Salas_Sokal_AT} for the 
direct SW-type 
algorithm \cite{Wiseman_Domany} for the Ashkin-Teller (AT) model  
\cite{Ashkin_Teller,Baxter}.

The important question is the following: Is the Li--Sokal bound 
\reff{Li_Sokal_bound_CH}/\reff{Li_Sokal_bound} sharp or not? 
An affirmative answer would imply that we could use the bound to predict 
the dynamic critical exponent(s) $z$ given only the static 
critical exponents of the system. 
There are three possibilities: 

\begin{enumerate} 
\item[i)] 
The bound \reff{Li_Sokal_bound_CH} is {\em sharp}\/ 
(i.e., the ratio $\tau/C_H$ is bounded), so that \reff{Li_Sokal_bound} is an  
{\em equality}\/. 

\item[ii)] 
The bound is {\em sharp modulo a logarithm}\/ 
(i.e., $\tau/C_H \sim \log^p L$ for $p>0$). 

\item[iii)] 
The bound is {\em not sharp}\/ 
(i.e., $\tau/C_H \sim L^p$ for $p>0$), so that \reff{Li_Sokal_bound} is a  
{\em strict inequality}\/.      
\end{enumerate}

Unfortunately, the empirical situation, even for the simplest cases, is 
far from clear. Let us review the status of this problem for the 2D Potts 
models. For the Ising case, the numerical data 
\cite{Swendsen_87,Heermann_90,Baillie_91,Baillie_92}  
are consistent with a power-law behavior with $z_{{\rm int},{\cal E}}$ 
ranging from 
$0.35\pm 0.01$ \cite{Swendsen_87} to 
$0.25 \pm 0.01$ 
\cite{Baillie_91,Baillie_92}. However, the data are also consistent  
with a logarithmic 
behavior $z_{{\rm int},{\cal E}} = 0\times\log$ \cite{Heermann_90}.   
In \cite{Salas_Sokal_AT} we reanalyzed the high-precision 
data of Baillie and Coddington \cite{Baillie_91,Baillie_92}\footnote{  
  The exact value of the specific heat at finite $L$ was taken from the
  paper by Ferdinand and Fisher \cite{Ferdinand_Fisher}.
},  
and we found that 
the ratio $\tau_{\rm int,{\cal E}}/C_H$ 
behaves most likely as a pure power law $\sim L^p$ with $p=0.060\pm 0.004$ 
(statistical error only) 
or as a logarithm $\sim A + B \log L$. This means that the bound 
\reff{Li_Sokal_bound_CH} is either non-sharp by a small power, or else 
it is sharp 
modulo a logarithm (in the latter case, the leading term would be  
$\tau_{{\rm int},{\cal E}} \sim \log^2 L$).  
It is extremely difficult to distinguish between these two 
scenarios with lattice sizes up to only $L=512$.

The 3-state Potts model was first considered under this perspective in 
\cite{Li_Sokal}. 
The dynamic critical exponent was found to be  
$z_{{\rm int},{\cal E}}=0.55\pm 0.03$, which is significantly larger  
than the exact result $\alpha/\nu = {2 \over 5} = 0.4$ \cite{Alexander}. 
So, it {\em seemed}\/ that the bound \reff{Li_Sokal_bound}
was not sharp at all.

The 4-state Potts model is rather peculiar: 
the naive fit to the data, $z_{{\rm int},{\cal E}} = 0.89 \pm 0.05$ 
\cite{Li_Sokal},
is {\em smaller}\/ than the (exactly known) 
value of $\alpha/\nu = 1$ \cite{Domany_Riedel}.
The explanation of this paradox is that the true leading term
in the specific heat has a multiplicative logarithmic correction,
$C_H \sim L \log^{-3/2} L$  
\cite{Salas_Sokal_FSS,Nauenberg_Scalapino,Cardy80,Black_Emery}.  
Indeed, a naive power-law fit to the specific heat yielded 
$\alpha/\nu = 0.75 \pm 0.01$, consistent with the bound  
\reff{Li_Sokal_bound}. A high-precision study of this model was carried out 
in \cite{Salas_Sokal_LAT95,Salas_Sokal_AT}.\footnote{ 
   In \cite{Salas_Sokal_LAT95,Salas_Sokal_AT} the 
   ``embedding'' version of an SW-type algorithm for the 
   AT model was used. This algorithm reduces to the standard SW 
   algorithm at the Ising subspace, but not at the 4-state Potts subspace. 
   However, it was shown numerically that both algorithms belong (as expected) 
   to the same dynamic universality 
   class at the 4-state Potts subspace. 
}  
Naive power-law fits to the data showed that the bound \reff{Li_Sokal_bound} 
was satisfied: $z_{{\rm int},{\cal E}}=0.876\pm0.012$  
and $\alpha/\nu=0.768\pm 0.009$. On the 
other hand, the behavior of the ratio $\tau_{{\rm int},{\cal E}}/C_H$ was 
consistent only with two scenarios: a power law $\sim L^p$ with  
$p=0.118\pm0.012$ (statistical error only),  
or a logarithmic growth $\sim A + B \log L$. This means that 
\reff{Li_Sokal_bound_CH} fails to be sharp either by a small power or by 
a logarithm. In conclusion, there are two likely behaviors for the 
autocorrelation time: either $\tau_{{\rm int},{\cal E}} \sim L^q \log^{-3/2}L$ 
with $q\approx1.12$, or else $\tau_{{\rm int},{\cal E}} \sim L\log^{-1/2}L$.  
In both cases, we find multiplicative logarithmic corrections to the 
autocorrelation time, 
which make the numerical analysis extremely difficult.  
These two scenarios for the ratio $\tau_{{\rm int},{\cal E}}/C_H$ coincide 
with those obtained for the Ising case.

In summary, for the 2D Potts models with $q=2$ and $q=4$,  
the Li--Sokal bound {\em might}\/ be either 
sharp modulo a logarithm or else {\em not sharp}\/ with a very small power 
$p \equiv z - \alpha/\nu$  
($0.04 \ltapprox p \ltapprox 0.12$).  
Moreover, if we interpolate between these models by following the self-dual 
(critical) curve of the AT model \cite{Ashkin_Teller,Baxter}, we obtain 
\cite{Salas_Sokal_LAT95,Salas_Sokal_AT} the same two scenarios:  
the ratio $\tau_{{\rm int},{\cal E}}/C_H$ either grows like a power law with 
a very small power $p$ 
(which increases slightly as we move from the Ising model to the 4-state Potts 
model along the AT self-dual curve), 
or else grows like a logarithm. Thus, there is some kind of continuity 
along the AT self-dual curve for the dynamic critical behavior of 
the SW algorithm.\footnote{  
   A similar study was carried out in 
   \protect\cite{Wiseman_Domany} for the single-cluster version of 
   the algorithm,
   but the lattices were not very large ($L\leq 256$).  
}

There results have led us to reappraise  
the status of the Li--Sokal bound for the 2D 
3-state Potts model. In \cite{Li_Sokal} the bound was declared 
{\em not sharp}\/ on the basis of numerical evidence suggesting that 
$p \equiv z - \alpha/\nu = 0.15 \pm 0.03$.  
However, this value is not much larger 
than that obtained \cite{Salas_Sokal_LAT95,Salas_Sokal_AT} for the 4-state 
Potts model; this suggests that the data for the 3-state model might also 
be consistent with a logarithm.  
Indeed, a closer look at the results of \cite{Li_Sokal} concerning the 3-state 
model reveals that the lattices studied 
were not very large ($L\leq 256$), and the 
statistics were not very high (the number of iterations for the $L=256$ 
lattice was at most $25 \times 10^3 \tau_{{\rm int},{\cal E}}$).  
This motivated us to 
reconsider this model and extend the results of \cite{Li_Sokal} to 
larger lattices, with higher statistics.

This paper is organized as follows: 
Section~\ref{sec_setup} reviews the basics of the Swendsen--Wang algorithm 
for the Potts models, as well as the proof of the Li--Sokal
bound for these models. In Section~\ref{sec_simul} we describe our 
Monte Carlo simulations. In Section~\ref{sec_analysis} we discuss in detail 
our methods of statistical data analysis. 
Finally, in Section~\ref{sec_res} we present the analysis of our numerical 
(static and dynamic) data, culminating in a discussion of the sharpness of 
the Li--Sokal bound.

\section{Basic set-up and notation} \label{sec_setup}

\subsection{Potts model and Swendsen--Wang algorithm} \label{sec_setup_notation}

The $q$-state Potts model assigns to each lattice site $i$ a spin variable 
$\sigma_i$ taking values in the set $\{1,2,\ldots,q\}$; 
these spins interact through the reduced Hamiltonian  
\be 
{\cal H}_{\rm Potts}   \;=\;   - \beta \sum_{\< ij \>} 
      (\delta_{\sigma_i,\sigma_j} - 1) \; ,  
\label{Potts_Hamiltonian} 
\ee  
where the sum runs over all the nearest-neighbor pairs $\<ij\>$. To simplify 
the notation we shall henceforth write  
$\delta_{\sigma_i,\sigma_j} \equiv \delta_{\sigma_b}$ for a bond $b=\<ij\>$.
The ferromagnetic case corresponds to $\beta>0$. The partition function is 
defined as 
\be 
Z   \;=\;   \sum_{\{\sigma\}} e^{-{\cal H}}   \;=\;   
    \sum_{\{\sigma\}}
       \exp\!\left[ \beta \sum\limits_b (\delta_{\sigma_b} - 1) \right]
    \;.
\label{Partition_function} 
\ee 
Finally, the Boltzmann weight of a configuration $\{\sigma\}$ is given by 
\be 
W_{\rm Potts}(\{\sigma\})   \;=\;
 {1 \over Z} \prod_b e^{\beta (\delta_{\sigma_b} - 1)}  
   \;=\;   {1 \over Z} \prod_b ( 1-p + p \delta_{\sigma_b} ) 
\label{Potts_weight} 
\ee
where $p=1-e^{-\beta}$.

The idea behind the Swendsen--Wang algorithm 
\cite{Sokal_Lausanne,Sokal_LAT90,Edwards_Sokal} is to decompose the  
Boltzmann weight by introducing 
new dynamical variables $n_b=0,1$ (living on the bonds of the lattice), 
and to simulate the joint model of old and new variables 
by alternately updating one set of variables conditional on the other set. 
The Boltzmann weight of the joint model is 
\be 
W_{\rm joint}(\{\sigma\},\{n\})   \;=\;   {1 \over Z} \prod_b \left[ 
   (1-p) \delta_{n_b,0} + p \delta_{\sigma_b} \delta_{n_b,1} \right] 
   \;.
\label{joint_weight} 
\ee 
The marginal distribution of 
\reff{joint_weight} with respect to the spin variables reproduces 
the Potts-model Boltzmann weight \reff{Potts_weight}. 
The marginal distribution of \reff{joint_weight} with 
respect to the bond variables is the 
Fortuin--Kasteleyn \cite{Kasteleyn_69,Fortuin_Kasteleyn_72,Fortuin_72} 
random-cluster model with parameter $q$:  
\be 
W_{\rm RC}(\{n\})   \;=\; 
   {1 \over Z} \left[ \prod_{b \colon\; n_b=1} p \right] 
 \left[ \prod_{b \colon\; n_b=0} (1-p) \right] q^{{\cal C}(\{n\})} 
\label{RC_weight} 
\ee 
where ${\cal C}(\{n\})$ is the number of connected components (including  
one-site components) in the graph whose edges are the bonds with $n_b=1$.

We can also consider the conditional probabilities of the joint 
distribution \reff{joint_weight}. 
The conditional distribution of the $\{n\}$ given the 
$\{\sigma\}$ is as follows: Independently for each bond $b = \<ij\>$, one sets 
$n_b=0$ when $\sigma_i \neq \sigma_j$, and sets 
$n_b=0$ and 1 with probabilities $1-p$ and $p$ when $\sigma_i=\sigma_j$. 
Finally, the conditional distribution of the $\{\sigma\}$ given the 
$\{n\}$ is as follows: Independently for each connected cluster, one sets 
all the spins $\sigma_i$ in that cluster equal to the same value, chosen 
with uniform probability from the set $\{1,2,\ldots,q\}$.   

The Swendsen--Wang algorithm simulates the
joint probability distribution \reff{joint_weight}
by alternately applying the two conditional distributions just described.
That is, we first erase the current $\{n\}$ configuration,
and generate a new $\{n\}$ configuration from the
conditional distribution given $\{\sigma\}$;
we then erase the current $\{\sigma\}$ configuration, 
and generate a new $\{\sigma\}$ configuration from the 
conditional distribution given $\{n\}$.

\subsection{Li--Sokal bound} \label{sec_setup_lisokal}

To prove the Li--Sokal bound we first notice that the transition matrix
$P_{SW}$ of the Swendsen--Wang algorithm can be written as a product
\be 
  P_{SW} \;=\;  P_{\rm bond} \, P_{\rm spin}   \;,
\ee 
where $P_{\rm bond}$ (the update of the bond variables)
and $P_{\rm spin}$ (the update of the spin variables)
are given by the conditional expectation operators
$E( \,\cdot\, | \{ \sigma\})$ and $E( \,\cdot\, | \{n\})$,
respectively.

The strategy behind the proof is to compute explicitly the autocorrelation 
function at time lags 0 and 1 for a suitable observable ${\cal O}$ 
(which we will choose to be a ``slow mode'' of the algorithm).  
The unnormalized autocorrelation function is defined as 
\be 
  C_{\cal OO}(t) \;\equiv\; \< {\cal O}(s) {\cal O}(s+t) \> - \<{\cal O}\>^2 
\label{def_c} 
\ee
(where the expectations are taken {\em in equilibrium}\/),
and the normalized autocorrelation function as 
\be
  \rho_{\cal OO}(t) \;\equiv\; {C_{\cal OO}(t) \over C_{\cal OO}(0) }  \;=\; 
                           {C_{\cal OO}(t) \over \var({\cal O}) }  
  \;.
\label{def_rho}
\ee
Then, using some
general properties of reversible Markov chains, we will deduce lower
bounds for the autocorrelation times $\tau_{{\rm int},{\cal O}}$  
and $\tau_{{\rm exp},{\cal O}}$.
These will in turn imply lower bounds on the
dynamic critical exponents $z_{{\rm int},{\cal O}}$ and   
$z_{{\rm exp},{\cal O}}$.

For the observable ${\cal O}$, we shall use the bond occupation
\be 
 {\cal N} \;\equiv\; \sum_b n_b  \;.
\label{def_n} 
\ee 
{}From the joint Boltzmann weight \reff{joint_weight} it is easy to compute 
the following bond expectation values conditional on the spin configuration 
$\{\sigma\}$: 
\begin{subeqnarray} 
E(n_b | \{\sigma\} )  &=& p \delta_{\sigma_b}  \\  
E(n_b n_{b'} | \{\sigma\} )  &=& \left\{ \begin{array}{ll} 
             p^2 \delta_{\sigma_b} \delta_{\sigma_{b'}} & 
             \quad \hbox{for $b\neq b'$}  \\ 
             p \delta_{\sigma_b}                        & 
             \quad \hbox{for $b=b'$}   
             \end{array} \right. 
\label{cond_n} 
\end{subeqnarray} 
{}From these equations it is easy to compute the mean values $\<{\cal N}\>$
and $\<{\cal N}^2\>$, and thus
$C_{\cal NN}(0) \equiv \var({\cal N}) \equiv \<{\cal N}^2\> - \<{\cal N}\>^2$:
\begin{subeqnarray} 
\<{\cal N}\>  &=& p \< {\cal E} \>   \slabel{eq2.11a}  \\ 
\<{\cal N}^2\>  &=& p^2 \< {\cal E}^2 \>  + p(1-p) \< {\cal E} \>  \\  
C_{\cal NN}(0) \equiv 
\var({\cal N})  &=& p^2 \var({\cal E}) + p(1-p) \< {\cal E} \> 
\label{cond_N} 
\end{subeqnarray} 
where the energy is defined as 
\be 
{\cal E} \;\equiv\; \sum_b \delta_{\sigma_b} \;.  
\label{def_energy} 
\ee 

The unnormalized autocorrelation function at time lag 1 is given by 
\be 
C_{\cal NN}(1) \;\equiv\; \< {\cal N}(0){\cal N}(1) \> - \< {\cal N} \>^2  
   \;=\; \var(P_{\rm bond}{\cal N}) \;=\; \var(E({\cal N}|\{\sigma\})) \;.  
\ee 
Now, $P_{\rm bond}{\cal N}$ is equal to 
\be 
P_{\rm bond}{\cal N} \;\equiv\; E({\cal N}|\{\sigma\}) \;=\;  
          p \sum_b \delta_{\sigma_b} \;=\; p {\cal E}   \;.  
  \label{eq2.14}
\ee
Therefore,
\be 
C_{\cal NN}(1) \;=\; p^2 \var({\cal E}) 
\ee
and  
\be 
\rho_{\cal NN}(1) \;\equiv\; {C_{\cal NN}(1) \over C_{\cal NN}(0)} \;=\;  
                    {p C_H \over p C_H + (1-p) E} \;=\; 1 - 
                    {(1-p) E \over p C_H + (1-p) E } 
\;,
\label{check_rho} 
\ee 
where the energy density $E$ and the specific heat $C_H$ have been defined as
\begin{eqnarray} 
E    &=& {1 \over V} \< {\cal E} \>
   \label{def_energy_density} \\ 
C_H  &=& {1 \over V} \var({\cal E})  \equiv {1\over V} \left[ 
         \< {\cal E}^2 \> - \< {\cal E} \>^2 \right] \;,  
   \label{def_cv}  
\end{eqnarray}
and $V$ is the number of lattice sites.

The correlation functions of ${\cal N}$ under
$P_{SW} \equiv P_{\rm bond}P_{\rm spin}$ 
are the same as under the positive-semidefinite self-adjoint operator
$P'_{SW} \equiv P_{\rm spin}P_{\rm bond}P_{\rm spin}$. This implies 
(see e.g., \cite{Sokal_Lausanne}) that
we have a spectral representation
\be
\rho_{\cal NN}(t) \;=\; \int_0^1 \lambda^{|t|} \, d\nu(\lambda)
  \label{spectral_rep}
\ee
with a positive measure $d\nu$. From this spectral representation 
we conclude that
\be
\rho_{\cal NN}(t) \;\geq\; \rho_{\cal NN}(1)^{|t|} \; .
\label{inequal1} 
\ee

If we now recall the definitions of the integrated and exponential 
autocorrelation times 
\begin{eqnarray} 
\tau_{{\rm int},{\cal N}}  &=& {1 \over 2} \sum_{t=-\infty}^{\infty} 
   \rho_{\cal NN}(t) 
  \label{def_tau_int} \\ 
\tau_{{\rm exp},{\cal N}}  &=& \lim_{|t|\rightarrow\infty} 
   {-|t| \over \log \rho_{\cal NN}(t) } 
  \label{def_tau_exp} 
\end{eqnarray} 
we conclude from \reff{check_rho}/\reff{inequal1} that 
\begin{eqnarray}
\tau_{{\rm int},{\cal N}}  &\geq& {1 \over 2} \times  
    {1 + \rho_{\cal NN}(1) \over 
     1 - \rho_{\cal NN}(1) } \geq {\rm const} \times C_H \\  
\tau_{{\rm exp},{\cal N}}  &\geq& {-1 \over \log \rho_{\cal NN}(1) }  
 \geq {\rm const} \times C_H  
\end{eqnarray}
These are precisely the bounds \reff{Li_Sokal_bound_CH}. If we take into 
account the expected behavior close to the critical point of the specific heat 
and the autocorrelation times, we deduce immediately the bounds 
\reff{Li_Sokal_bound}.

Similar bounds hold for the autocorrelation times of the 
energy ${\cal E}$. This can be 
seen from the fact that $P_{\rm bond}{\cal N} = p {\cal E}$,
which implies that
\be
   C_{\cal EE}(t)  \;=\; p^{-2} C_{\cal NN}(t+1)
  \label{check_c_1}
\ee
and hence
\be 
\rho_{\cal EE}(t) \;=\; {\rho_{\cal NN}(t+1) \over \rho_{\cal NN}(1) } \;\geq\; 
\rho_{\cal NN}(t) \;.  
\label{inequal2} 
\ee
This equation proves that 
\be 
\tau_{{\rm exp},{\cal E}} \;=\; \tau_{{\rm exp},{\cal N}}    \;,
\label{equal2} 
\ee 
and furthermore allows us to bound $\tau_{{\rm int},{\cal E}}$
above and below in terms of $\tau_{{\rm int},{\cal N}}$:  
\be 
\tau_{{\rm int},{\cal N}} \;\leq\; \tau_{{\rm int},{\cal E}} 
\;=\; { \tau_{{\rm int},{\cal N}} - {1 \over 2}
    \over
    \rho_{\cal NN}(1)
  }
  - {1 \over 2} \;\leq\;  
{ \tau_{{\rm int},{\cal N}} \over \rho_{\cal NN}(1) } \;.  
\label{bounds_tau_e} 
\ee
If the critical slowing-down is not completely eliminated, we 
expect the factor $\rho_{\cal NN}(1)$ to approach 1 (from 
below) as $L\rightarrow\infty$.
Moreover, irrespective of the presence or absence 
of critical slowing-down, we expect $\rho_{\cal NN}(1)$ to be 
{\em bounded away from zero}\/ as $L\rightarrow\infty$.
Modulo this very weak hypothesis,
\reff{bounds_tau_e} implies the equality of the dynamic 
critical exponents for the energy and bond-occupation observables:
\be
z_{{\rm int},{\cal E}} \;=\; z_{{\rm int},{\cal N}}   \;.
\label{equal3}
\ee
Unfortunately, we do not know how to {\em rigorously}\/ rule out ``exotic'' 
behaviors, in which $\rho_{\cal NN}(1)$ tends to {\em zero}\/ as 
$L\rightarrow\infty$ and yet $\tau_{{\rm int},{\cal N}}$ diverges because  
the autocorrelation function has an extremely long tail.

Finally, we can define a new (``energy-like'') bond observable:
the nearest-neighbor connectivity  
\be 
{\cal E}' \;\equiv\; \sum_b \gamma_b \;,
\label{def_energyp} 
\ee 
where $\gamma_b$ equals 1 if both ends of the bond $b$ belong to the
same cluster, and 0 otherwise.
More generally, the connectivity $\gamma_{ij}$ can be defined
for an arbitrary pair $i,j$ of sites:
\be
  \gamma_{ij}(\{n\}) \;=\; \left\{ \begin{array}{ll}
       1 & \quad \hbox{\rm if $i$ is connected to $j$} \\
       0 & \quad \hbox{\rm if $i$ is not connected to $j$}
       \end{array} \right.
\ee
The interest in ${\cal E}'$ stems from the fact that the
conditional expectation of $\delta_{\sigma_b}$ given 
the bond configuration $\{n\}$ is essentially $\gamma_b$:
\be
E(1-\delta_{\sigma_b}|\{n\}) \;=\; {q -1 \over q} (1 - \gamma_b )   \;.
\label{cond_energy} 
\ee
This implies the following relation between 
the energy and 
the connectivity densities:
\be 
{k \over 2}  - E \;=\; {q-1 \over q} \left({k \over 2} - E' \right) \;,  
\label{check_enerp} 
\ee 
where $k$ is the coordination number of the lattice  
(i.e., the number of nearest neighbors of any given site), 
and the connectivity density $E'$ is defined as 
\be
   E' \;=\; {1 \over V} \< {\cal E}' \> \;. 
\label{def_energyp_density}
\ee
Furthermore, \reff{cond_energy} tells us that 
\be 
P_{\rm spin}(kV/2 - {\cal E})  \;=\;
 P_{\rm spin}\sum_b (1 - \delta_{\sigma_b})  \;=\;
   {q-1 \over q} \sum_b(1 - \gamma_b)  \;=\;
   {q-1 \over q} (kV/2 - {\cal E}')  \;.  
  \label{eq2.35}
\ee 
This implies that 
\be
   C_{\cal E'E'}(t) \;=\;
      \left( {q \over q-1} \right)^{\! 2}  C_{\cal EE}(t+1)
 \label{check_c_2}
\ee
and hence
\be
\rho_{\cal E'E'}(t) \;=\;
 {\rho_{\cal EE}(t+1) \over \rho_{\cal EE}(1) } \;\geq\;
\rho_{\cal EE}(t) \;.
\label{inequal3}
\ee
Thus, the bounds 
\reff{Li_Sokal_bound_CH}/\reff{Li_Sokal_bound} hold also for the 
autocorrelation times of ${\cal E}'$. Furthermore, we can obtain 
bounds analogous to \reff{equal2}--\reff{equal3} for the observable 
${\cal E}'$:   
\be  
\tau_{{\rm exp},{\cal E}} \;=\; \tau_{{\rm exp},{\cal E}'}
\ee
and 
\be 
\tau_{{\rm int},{\cal E}} \;\leq\; \tau_{{\rm int},{\cal E}'}
\;\leq\; { \tau_{{\rm int},{\cal E}} \over \rho_{\cal EE}(1) } \;.  
\label{bounds_tau_ep} 
\ee  
Using again \reff{bounds_tau_e} and \reff{inequal2} we arrive at 
\be
\tau_{{\rm int},{\cal N}} \;\leq\; \tau_{{\rm int},{\cal E}'}
\;\leq\; { \tau_{{\rm int},{\cal N}} \over \rho_{\cal NN}(2) } \;.
\label{bounds_tau_ep2} 
\ee
If $\rho_{\cal NN}(2)$ is bounded away from zero as $L\rightarrow\infty$, 
we can conclude that the dynamic critical exponents of
${\cal N}$, ${\cal E}$ and ${\cal E}'$ are all equal:  
\be
z_{{\rm int},{\cal N}}   \;=\;
z_{{\rm int},{\cal E}}  \;=\;  z_{{\rm int},{\cal E}'}    \;.
  \label{equal4}
\ee

\section{Description of the simulations} \label{sec_simul}

\subsection{Observables to be measured} \label{sec_obs}

We have made simulations of the two-dimensional 
3-state Potts model at criticality,  
\be 
  \beta \;=\; \beta_c \;=\; \log (1 + \sqrt{3}) \;\approx\; 1.00505254 \;,  
\ee 
on a periodic square lattice of linear size $L$.

We have measured five basic observables. Three of them have been 
already defined: the energy ${\cal E}$ \reff{def_energy},  
the bond occupation  ${\cal N}$ \reff{def_n}, and the 
nearest-neighbor connectivity ${\cal E}'$ \reff{def_energyp}.
The other two are 
\begin{subeqnarray} 
{\cal M}^2  &=& \left( \sum_x \bsigma_x \right)^2        \\
            &=& {q \over q-1} \sum_{\alpha=1}^q \left(
     \sum_x \delta_{\sigma_x,\alpha} \right)^2 - {V^2 \over q-1}  
 \label{def_msquare}
\end{subeqnarray} 
and
\begin{subeqnarray}
{\cal F}    &=&
{1 \over 2} \left( \left| \sum_x \bsigma_x \, e^{2\pi i x_1/L} \right|^2   +
                   \left| \sum_x \bsigma_x \, e^{2\pi i x_2/L} \right|^2
            \right)
     \\
  &=& {q \over q-1} \times {1 \over 2} \sum_{\alpha=1}^q \left(
   \left| \sum_x \delta_{\sigma_x,\alpha} \, e^{2\pi i x_1/L} \right|^2 + 
   \left| \sum_x \delta_{\sigma_x,\alpha} \, e^{2\pi i x_2/L} \right|^2 \right)
 \label{def_f} 
\end{subeqnarray} 
where $\bsigma_x \in \R^{q-1}$ is the Potts spin in the
hypertetrahedral representation
[for $q=3$ this means
 $\bsigma_x = (\cos (2\pi \sigma_x/3), \sin (2\pi \sigma_x/3))$],
$V=L^2$ is the number of lattice sites,
and $(x_1,x_2)$  are the Cartesian coordinates of point $x$.
The observable ${\cal F}$ can be regarded as the square 
of the Fourier transform of the spin variable at the smallest allowed non-zero 
momenta [i.e., $(\pm2\pi/L,0)$ and $(0,\pm2\pi/L)$ for the square lattice]. 
It is normalized to be comparable to its zero-momentum analogue ${\cal M}^2$.

{}From these observables we compute the following expectation values: 
the energy density $E$ \reff{def_energy_density},  
the specific heat $C_H$ \reff{def_cv}, the connectivity density $E'$ 
\reff{def_energyp_density}, the 
bond density 
\be 
   N \;=\; {1 \over V} \< {\cal N} \>  \;,  
\label{def_n_density} 
\ee 
the magnetic susceptibility
\be 
  \chi \;=\; {1 \over V} \< {\cal M}^2 \> \;,  
\label{def_susceptibility} 
\ee 
the correlation function at momentum $(2\pi/L,0)$
\be 
 F \;=\; {1 \over V} \< {\cal F} \>  \;,
 \label{def_f_density} 
\ee 
and the second-moment correlation length 
\be 
\xi \;=\; { \left( {\chi \over F} - 1 \right)^{1/2} 
        \over 
        2 \sin (\pi/L)} \;.
\label{def_xi} 
\ee 
This definition of the correlation length is {\em not}\/ equal
to the exponential 
correlation length (=1/mass gap); but it is expected that both correlation 
lengths scale in the same way as we approach the critical point. 

\medskip 

\noindent 
{\bf Remark:} As a check we have also computed the mean-square size of the 
clusters, 
\be 
{\cal S}_2 \;=\; \sum_{\cal C} \#({\cal C})^2 \;,  
\label{def_s2}  
\ee 
where the sum is over all the clusters ${\cal C}$ of activated bonds  
(i.e., with $n_b=1$) and $\#({\cal C})$ is the number of sites of the cluster 
${\cal C}$. Using the Fortuin--Kasteleyn identities 
\cite{Sokal_Lausanne,Kasteleyn_69,Fortuin_Kasteleyn_72,Fortuin_72}, 
it is not difficult to show that 
\be 
\< {\cal S}_2 \> \;=\; \< {\cal M}^2 \> \;.  
\label{check_stwo} 
\ee 

\medskip 

For each observable ${\cal O}$ discussed above we have measured its 
autocorrelation function $\rho_{\cal OO}(t)$ \reff{def_rho}; and from this 
function we have estimated the corresponding integrated autocorrelation time 
$\tau_{{\rm int},{\cal O}}$ \reff{def_tau_int}. In 
Section~\ref{sec_analysis} we explain in detail how we derived estimates  
of the mean values and the error bars for both static and 
dynamic quantities.

\subsection{Summary of the simulations} \label{sec_summary}

We have run our Monte Carlo  
program on lattices with $L$ ranging from 4 to 1024. 
In all cases the initial configuration was random, and we have discarded 
the first $10^5$ iterations to allow the 
system to reach equilibrium; this discard interval is  
always greater than $10^3 \, \tau_{{\rm int},{\cal E}}$.\footnote{ 
   We expect that  
   $\tau_{{\rm int},{\cal E}}/\tau_{{\rm exp},{\cal E}} \approx 0.96$
   and $\tau_{{\rm exp},{\cal E}} = \tau_{{\rm exp}}$
   for this 
   algorithm \cite{Salas_Sokal_AT} (see Section~\protect\ref{sec_res_tauexp}). 
   So the discard interval is always greater than
   $10^3 \tau_{{\rm exp}}$, which is more than sufficient.
} 
For $4 \leq L \leq 256$ the total run length was approximately 
$10^6 \, \tau_{{\rm int},{\cal E}}$;  
for $L=512$, it was $ 2.2 \times 10^5 \, \tau_{{\rm int},{\cal E}}$; 
and for $L=1024$, it was $ 6.7 \times 10^4 \, \tau_{{\rm int},{\cal E}}$. 
In all 
cases, the statistics are high enough to permit a high accuracy in our 
estimates of the static (error $\sim$ 0.1--0.5\%) 
and dynamic (error $\sim$ 0.5--2\%) 
quantities.

For $4\leq L\leq 128$ our data were obtained from a single long run. 
For $L=256$ we made two independent runs (with different random-number-generator
seeds); for $L=512$ we made three independent runs;  
and for $L=1024$ we made four independent runs. In each case, we discarded 
the first $10^5$ iterations of each run. The individual runs are all of length  
$\gtapprox 10^4 \tau_{{\rm int},{\cal E}}$, which is long enough to allow  
a good determination of the dynamic quantities.

To test the program, we compared the static results on the $4\times4$ lattice 
to the exact solution (obtained by enumerating all the possible 
configurations on this lattice). The agreement was excellent (see 
Table~\ref{table_static_full}). In addition, for all lattice sizes we  
checked the relations \reff{eq2.11a}, \reff{check_enerp} and
\reff{check_stwo} between the mean 
values of static observables; the relations \reff{check_c_1}/\reff{check_c_2}
among  the autocorrelation functions $C_{\cal NN}$, $C_{\cal EE}$
and $C_{\cal E'E'}$;
and the relation \reff{check_rho} between the autocorrelation 
function $\rho_{\cal NN}(1)$ and the static observables $E$ and $C_H$. In all 
cases the agreement was also good.

The CPU time required by our program is approximately  
7.2 $L^2$ $\mu$s/iteration on an IBM RS-6000/370, and  
3.6 $L^2$ $\mu$s/iteration on a single processor of an IBM RS/6000 SP2.  
The total CPU time used in 
the project was approximately 1.5 years on an IBM RS/6000 SP2. 
The smallest lattices were run on a IBM RS-6000/370 at NYU, while the 
largest ones ($L\geq 128$) were run on the IBM SP2 cluster at the 
Cornell Theory Center.

\section{Statistical analysis of the Monte Carlo data} \label{sec_analysis} 

In this paper we are aiming at extremely high precision
for both static and dynamic quantities;
and furthermore we need to disentangle the effects of statistical errors
from the effects of systematic errors due to corrections to scaling.
For this, it is essential to obtain accurate estimates
not only of the static and dynamic quantities of interest,
but also of their {\em error bars}\/:
in this way we will be able (see Section \ref{sec_res})
to perform $\chi^2$ tests which provide an objective measure
of the goodness of fit in each scaling Ansatz.

In this section we discuss in some detail how we performed the  
statistical analysis of our raw Monte Carlo data.
In particular, we describe how to compute the estimators 
for the mean value and the variance of both static and dynamic 
quantities. These methods are based on well-known  
results of time-series analysis \cite{Anderson,Priestley},
which we review briefly in Section \ref {sec_analysis_standard}.\footnote{ 
   A review of time-series-analysis methods as applied to MC simulations  
   can be found 
   in \cite[Appendix C]{Madras_Sokal} and in \cite{Sokal_Lausanne}. 
}  
Then (Section \ref{sec_analysis_bunches})
we describe an alternative analysis method,
based on independent ``bunches'',
and report the results of detailed cross-checks that confirm
(with one slight exception) the validity and reliability
of the standard time-series-analysis method.

\subsection{``Standard'' time-series-analysis method}  
\label{sec_analysis_standard}

Let us consider a generic observable ${\cal O}$, whose mean is equal to 
$\mu_{\cal O}$. Its corresponding unnormalized and normalized autocorrelation 
functions are denoted by 
$C_{\cal OO}(t) \equiv \<{\cal O}(0){\cal O}(t)\> - \<{\cal O}\>^2$ and 
$\rho_{\cal OO}(t) \equiv C_{\cal OO}(t)/C_{\cal OO}(0)$, 
respectively. 
We also define the integrated autocorrelation time  
\be 
\tau_{{\rm int},{\cal O}} \;=\; {1\over 2}  
                    \sum_{t=-\infty}^\infty \rho_{\cal OO}(t)  \;.
\ee 

Given a sequence of $n$ Monte Carlo measurements 
of the observable ${\cal O}$ --- call them 
$\{ {\cal O}_1, \ldots, {\cal O}_n \}$ --- the natural estimator of the 
mean $\mu_{\cal O}$ is the sample mean  
\be
\overline{{\cal O}}  \;\equiv\;  {1 \over n} \sum_{i=1}^n {\cal O}_i \;. 
\ee
This estimator is unbiased and has a variance
\begin{subeqnarray}
{\rm var}(\overline{{\cal O}}) &=& {1 \over n^2}
                            \sum_{r,s=1}^n C_{\cal OO}(r-s)
    \\
                              &=& {1 \over n} \sum_{t=-(n-1)}^{n-1}
    \left( 1 - {|t| \over n} \right) C_{\cal OO}(t)  \\
                              &\approx& {1 \over n}\
                              2 \tau_{ {\rm int},{\cal O} } \
                               C_{\cal OO}(0) \quad \mbox{for} \quad
    n \gg \tau_{ {\rm int},{\cal O} }
\label{var_mean} 
\end{subeqnarray}
This means that the variance is a factor $2 \tau_{ {\rm int},{\cal O} }$
larger than it would be if the measurements were uncorrelated. It is, therefore,
very important to estimate the autocorrelation time $\tau_{{\rm int},{\cal O}}$ 
in order to ensure a correct determination of the error bar on  
the (static) quantity $\mu_{\cal O}$.

The natural estimator for the unnormalized autocorrelation function 
$C_{\cal OO}(t)$ is 
\be 
\widehat{C}_{\cal OO}(t) \;\equiv\; {1 \over n - |t|} \sum_{i=1}^{n-|t|} 
           ({\cal O}_i - \mu_{\cal O})({\cal O}_{i+|t|} - \mu_{\cal O}) 
\ee 
if the mean $\mu_{\cal O}$ is known, and   
\be
\widehat{\widehat{C}}_{\cal OO}(t) 
           \;\equiv\; {1 \over n - |t|} \sum_{i=1}^{n-|t|}
           ({\cal O}_i - \overline{\cal O}) 
           ({\cal O}_{i+|t|} - \overline{\cal O})  
\ee
if the mean $\mu_{\cal O}$ is unknown. 
We emphasize that, for each $t$, the estimators  
$\widehat{C}_{\cal OO}(t)$ and
$\widehat{\widehat{C}}_{\cal OO}(t)$ are
{\em random variables}\/ [in contrast to $C_{\cal OO}(t)$, which is a 
{\em number}\/]. 
The estimator $\widehat{C}_{\cal OO}(t)$ is unbiased, and  
$\widehat{\widehat{C}}_{\cal OO}(t)$ is biased by terms of 
order $1/n$. 
The covariance matrices of $\widehat{C}_{\cal OO}$ and 
$\widehat{\widehat{C}}_{\cal OO}$ are the same to leading order  
in the large-$n$ limit 
(i.e., $n \gg \tau_{{\rm int},{\cal O}}$), and we have 
\cite{Anderson,Priestley}  
\begin{eqnarray}  
{\rm cov}(\widehat{C}_{\cal OO}(t),\widehat{C}_{\cal OO}(u)) &=& 
{1 \over n} \sum_{i=-\infty}^{\infty} \left[ C_{\cal OO}(m) C_{\cal OO}(m+u-t) 
 +C_{\cal OO}(m+u)C_{\cal OO}(m-t) \right. \nonumber \\ 
 & & \qquad \qquad \left. + \kappa(t,m,m+u)\right] + o\left({1\over n}\right) 
 \;, 
\label{cov_c} 
\end{eqnarray} 
where $t,u \geq 0$ and $\kappa$ is the connected 4-point autocorrelation  
function   
\begin{eqnarray}  
\kappa(r,s,t) &\equiv& 
        \< ({\cal O}_i -\mu_{\cal O})    ({\cal O}_{i+r} -\mu_{\cal O})
           ({\cal O}_{i+s} -\mu_{\cal O})({\cal O}_{i+t} -\mu_{\cal O})\>   
\nonumber \\ 
 & & \qquad - C_{\cal OO}(r)C_{\cal OO}(t-s) - C_{\cal OO}(s)C_{\cal OO}(t-r) 
 \nonumber \\ 
 & & \qquad 
     - C_{\cal OO}(t)C_{\cal OO}(s-r) \;. 
\end{eqnarray} 

The natural estimator for the normalized autocorrelation function 
$\rho_{\cal OO}(t)$ is 
\be 
\widehat{\rho}_{\cal OO}(t) \;\equiv\; {\widehat{C}_{\cal OO}(t) \over 
                                  \widehat{C}_{\cal OO}(0) }  
\label{estim1} 
\ee 
if the mean $\mu_{\cal O}$ is known, and 
\be
\widehat{\widehat{\rho}}_{\cal OO}(t) \;\equiv\; 
        {\widehat{\widehat{C}}_{\cal OO}(t) \over
         \widehat{\widehat{C}}_{\cal OO}(0) } 
\label{estim2} 
\ee
if the mean $\mu_{\cal O}$ is unknown. 
The estimators $\widehat{\rho}_{\cal OO}(t)$ and 
$\widehat{\widehat{\rho}}_{\cal OO}(t)$ are biased by terms of order $1/n$,  
as a result of the ratios of random variables in   
\reff{estim1}/\reff{estim2}. 
The covariance matrices of $\widehat{\rho}_{\cal OO}$ and 
$\widehat{\widehat{\rho}}_{\cal OO}$ are the same to leading order in 
$1/n$. 
If the process is Gaussian,
this covariance matrix is given in the large-$n$ limit by 
\cite{Priestley} 
\begin{eqnarray}  
{\rm cov}(\widehat{\rho}_{\cal OO}(t),\widehat{\rho}_{\cal OO}(u)) &=&    
{1 \over n} \sum_{m=-\infty}^{\infty} \left[  
  \rho_{\cal OO}(m)\rho_{\cal OO}(m+t-u) + 
  \rho_{\cal OO}(m+u)\rho_{\cal OO}(m-t)  \right. \nonumber \\ 
 & & + 2\rho_{\cal OO}(t)\rho_{\cal OO}(u)\rho_{\cal OO}^2(m) 
     - 2\rho_{\cal OO}(t)\rho_{\cal OO}(m)\rho_{\cal OO}(m-u) \nonumber \\ 
 & & \left.  
     - 2\rho_{\cal OO}(u)\rho_{\cal OO}(m)\rho_{\cal OO}(m-t) \right] + 
     o\left({1 \over n}\right) \;. 
\label{cov_rho} 
\end{eqnarray} 
If the process is {\em not}\/ Gaussian, then there are additional terms  
proportional to the fourth cumulant $\kappa(m,t,t-u)$; these terms are, like 
those in \reff{cov_rho}, of order $1/n$. 
The simplest assumption is to consider the stochastic process to be 
``not too far from Gaussian'', and drop all the terms involving $\kappa$.    
If this assumption is {\em not}\/ justified, then we are introducing a bias in 
the estimate of this covariance.

Finally, we shall take 
the estimator for the integrated autocorrelation time to be 
\cite{Madras_Sokal} 
\be 
\widehat{\tau}_{{\rm int},{\cal O}} \;\equiv\; {1\over 2} \sum_{t=-M}^M 
            \widehat{\rho}_{\cal OO}(t)  
\ee 
[or the same thing with $\widehat{\widehat{\rho}}_{\cal OO}(t)$]
where $M$ is a suitably chosen number. The reason behind the cutoff $M$ is 
the following: if we were to make the ``obvious'' choice $M=n+1$, then the 
resulting estimator would have a variance of order 1 even in the limit  
$n\rightarrow\infty$;  
this is because the terms $\widehat{\rho}_{\cal OO}(t)$ with 
large $t$ have errors (of order $1/n$) that {\em do not}\/ vanish as $t$ 
grows [cf. \reff{cov_rho}], and their number is 
also large ($\sim n$). 
Taking $M \ll n$ restores the good behavior of the estimator as 
$n\rightarrow\infty$. 
The bias 
introduced by this rectangular cutoff\footnote{   
   We could use more general cutoff functions, but this rectangular cutoff 
   is the most convenient for the present purposes.  
} 
is given by 
\be 
{\rm bias}(\widehat{\tau}_{{\rm int},{\cal O}}) \;=\; 
 - {1 \over 2} \sum_{|t| > M} \rho_{\cal OO}(t) + o\left({1\over n}\right)  
\;. 
\label{bias_tauint} 
\ee 
The variance of the estimator $\widehat{\tau}_{{\rm int},{\cal O}}$ can 
be computed from the covariance \reff{cov_rho};  
the final result is \cite{Madras_Sokal}  
\be 
{\rm var}(\widehat{\tau}_{{\rm int},{\cal O}}) \;\approx\; 
{2(2M + 1) \over n} \tau_{{\rm int},{\cal O}}^2 \;,  
\label{var_tauint} 
\ee
where the approximation $\tau_{{\rm int},{\cal O}} \ll M \ll n$ has been made. 
A good (self-consistent) choice of $M$ is the following \cite{Madras_Sokal}:
let $M$ be the smallest integer such that 
$M \geq c \widehat{\tau}_{{\rm int},{\cal O}}(M)$, where $c$ is a suitable 
constant. 
If the normalized autocorrelation function is roughly a pure 
exponential\footnote{ 
   This has been found empirically to be true in the SW algorithm for
   the 2D Ising \cite{Sokal_unpublished} and 4-state Potts models 
   \cite{Salas_Sokal_AT}, and is confirmed here for the 3-state Potts model 
   (see Section~\ref{sec_res_tauexp}). 
}, 
then the choice $c\approx 6$ is reasonable. Indeed, if we take 
$\rho_{\cal OO}(t) = e^{-t/\tau}$ and minimize the 
mean-square error 
\be 
\hbox{MSE}(\widehat{\tau}_{{\rm int},{\cal O}}) \;\equiv\; 
{\rm bias}(\widehat{\tau}_{{\rm int},{\cal O}})^2 + 
{\rm var}(\widehat{\tau}_{{\rm int},{\cal O}}) 
\ee 
using \reff{bias_tauint}/\reff{var_tauint}, we find that the optimal 
window width is
\be 
M_{\rm opt} \;=\; {\tau \over 2} \log \left( {n \over 2 \tau} \right) -1 \;. 
\ee
For $n/\tau \approx 10^6$ (resp. $10^4$), we have 
$M_{\rm opt}/\tau \approx 6.56$ (resp. 4.26).

As noted above, we expect the estimator  
$\widehat{\tau}_{{\rm int},{\cal O}}$ to have a bias of order 
$\tau_{{\rm int},{\cal O}}/n$, due to the nonlinearities in 
\reff{estim1}/\reff{estim2}.\footnote{   
    The bias on the estimator $\widehat{\tau}_{{\rm int},{\cal O}}$ also  
    induces a bias on the 
    estimated variance \reff{var_mean} 
    of the sample mean $\overline{\cal O}$.  
    This bias is of order $1/n^2$, i.e.\ a factor $1/n$ down from
    the variance \reff{var_mean} itself.
}  
To make this bias negligible we need long 
runs. It has been shown empirically that this procedure works fairly well  
when $n \gtapprox 10^4 \widehat{\tau}_{{\rm int},{\cal O}}$   
\cite{Sokal_Lausanne}.  

\medskip 

\noindent 
{\bf Remarks:} 1. The estimation of the error bar for the specific heat is 
a little bit more complicated. One can obtain var($C_H$) by computing 
var(${\cal E}$), var(${\cal E}^2$), and cov(${\cal E},{\cal E}^2$). 
This procedure has a numerical 
drawback: sometimes the covariance matrix for the observables ${\cal E}$ and 
${\cal E}^2$ is nearly singular; then a small statistical fluctuation can  
cause the {\em estimator}\/  of this  
matrix to be non-positive-definite. We are not aware of any  
procedure that ensures that the estimator of a covariance matrix is  
also positive-definite.  
To overcome this difficulty, we considered the observable 
${\cal O} = ({\cal E} - \mu_{\cal E})^2$, which can be studied using 
the standard method. As we do not know exactly the value of 
$\mu_{\cal E}$, we can use instead the sample mean 
$\overline{\cal E}$ (which should be computed first).  
To leading order in $1/n$ this procedure gives the right error bar for the 
specific heat. 

\medskip 

2. We have a similar problem when computing the error bar for the second-moment
correlation length $\xi$, which is a function of the two quantities 
$\chi$ and $F$ [cf. \reff{def_f_density}]. 
In this case we considered the random variable 
\be 
{\cal O}' \;=\; { {{\cal M}^2 \over \mu_{{\cal M}^2} } - 
              {{\cal F}   \over \mu_{\cal F}     } } \;,  
\ee
which has automatically a zero mean. The error bar for the 
second-moment correlation length can be written easily as a function of 
the susceptibility $\chi$, the quantity $F$, and the variance of the 
above-mentioned observable ${\cal O}'$. With this trick we take into account 
the cross-correlation between ${\cal M}^2$ and ${\cal F}$. This method 
needs the mean values $\mu_{{\cal M}^2}$ and $\mu_{\cal F}$;  
in practical situations they are substituted by the corresponding sample 
means $\overline{{\cal M}^2}$ and $\overline{\cal F}$ (which must 
be computed first).  
  
\medskip 
 
This is the standard procedure we have used to analyze each of our 
Monte Carlo runs. Estimates coming from multiple independent runs (for  
$L\geq 256$) were merged using the standard formulae for statistically 
independent data.  
The results are reported in Table~\ref{table_static_full} (static quantities) 
and Table~\ref{table_dynamic_full} (dynamic quantities). The results on 
$\tau_{{\rm int},{\cal E}}$ for $16 \leq L \leq 256$ can be compared 
with those reported in \cite{Li_Sokal}; the agreement is excellent
($\chi^2=6.17$, 5 degrees of freedom, confidence level = 56\%).

\subsection{Method based on independent bunches} 
\label{sec_analysis_bunches}

Unfortunately, this standard method does not provide an easy way to 
compute the error bar for complicated quantities such as the ratio 
$\tau_{{\rm int},{\cal E}}/C_H$, which will play a central role in our 
analysis of the sharpness of the Li--Sokal bound  
(see Section~\ref{sec_res_li_sokal}).    
We can, of course,  
give an {\em upper bound}\/  on the actual error bar by using the triangle  
inequality; but this upper bound will be a significant overestimate of the 
true value. If we were then to use this overestimate in fits, we would 
find artificially small values of $\chi^2$; as a result, the confidence levels 
would be artificially high, and useless for distinguishing good from bad fits.
(At best we could distinguish {\em better}\/ versus {\em worse}\/ fits, 
by looking at the {\em relative}\/ values of $\chi^2$.)

This fact motivated us to look for an alternative method to compute 
the error bars. There is also another advantage in having an alternate 
method: we can independently check  
the assumptions and approximations made in the standard procedure.

The second method works as follows. First, we split the whole sample of 
$n$ MC measurements $\{ {\cal O}_1, {\cal O}_2, \ldots, {\cal O}_n \}$ 
into $m$ bunches of equal (or almost equal) length $\ell \equiv n/m$.  
For each bunch $i$ ($1\leq i \leq m$)  
we can compute the sample mean of the observable ${\cal O}$ and 
an estimate of its 
integrated autocorrelation time ($\overline{\cal O}_i$ and 
$\widehat{\tau}_{{\rm int},{\cal O},i}$, respectively). Indeed, we can also 
estimate the corresponding variances, but these estimates 
do {\em not}\/ play any role in 
what follows. When computing the autocorrelation functions within each 
bunch, we used the whole sample mean $\overline{\cal O}$ as our estimate
of $\mu_{\cal O}$ (instead of the bunch sample mean
$\overline{\cal O}_i$); this trick reduces the bias in the estimates of 
the autocorrelation functions.

In this way we obtain two sequences of single-bunch estimates
$\{\overline{\cal O}_i \}$ and $\{\widehat{\tau}_{{\rm int},{\cal O},i} \}$. 
If the bunches are long enough (i.e. $\ell \gg \tau_{{\rm int},{\cal O}}$), 
then the estimates from distinct bunches are almost statistically independent.  
Thus, we can define our estimates as follows:
\begin{subeqnarray} 
\widehat{\cal O}' &\equiv& {1 \over m} \sum_{i=1}^m 
                          \overline{\cal O}_i    \\
\widehat{\rm var}(\widehat{\cal O}') &\equiv& {1 \over m(m-1)} \sum_{i=1}^m 
         (\overline{\cal O}_i - \widehat{\cal O}')^2
\slabel{O_error_bunch}  \\ 
\widehat{\tau}^\prime_{{\rm int},{\cal O}} &\equiv& 
           {1 \over m} \sum_{i=1}^m 
     \widehat{\tau}_{{\rm int},{\cal O},i}   
\slabel{tau_int_bunch} \\ 
\widehat{\rm var}(\widehat{\tau}^\prime_{{\rm int},{\cal O}}) &\equiv& 
      {1 \over m(m-1)} \sum_{i=1}^m 
  (\widehat{\tau}_{{\rm int},{\cal O},i} -  
   \widehat{\tau}^\prime_{{\rm int},{\cal O}})^2
\slabel{tau_int_error_bunch}
\end{subeqnarray} 
The quality of the results depends on the total number of 
measurements $n$ and the number $m$ of bunches we use. The merging of data 
coming from different runs is trivial in this case.

For $4 \leq L \leq 256$ we have extremely good statistics 
($n \sim 10^6 \, \tau_{{\rm int},{\cal E}}$). This allows us to vary 
$m$ over at least one order of magnitude, and thereby to provide a 
cross-check on the standard time-series analysis. 
In this discussion it is useful to divide the observables into three  
categories: linear static, non-linear static, and dynamic. 
The first category includes
${\cal E}$, ${\cal E}'$, ${\cal N}$, ${\cal M}^2$ and ${\cal F}$, whose
sample mean values are linear in the raw MC data.
The second category includes the specific heat and the second-moment 
correlation length, whose mean values are non-linear functions of the 
raw MC data. 
Finally, the third category contains all the autocorrelation times, which 
are also non-linear functions of the raw data.

For $4 \leq L \leq 256$ we first divided the whole sample into $m=100$ 
bunches, each of 
them with a length $\ell \sim 10^4 \, \tau_{{\rm int},{\cal E}}$.
%%  Thus, we 
%%  expect that the determination of both the static and dynamic quantities 
%%  within each bunch will be accurate. 
For $4 \leq L\leq 64$ we repeated the analysis using $m=1000$ bunches,
each of them with
a length $\ell \sim 10^3 \, \tau_{{\rm int},{\cal E}}$.

%
% 1st category 
%
For the linear static observables, the ``standard'' and ``bunch'' methods
always give identical mean values;
this is a trivial identity, provided that the bunch lengths  
are exactly equal.
As for the error bars on these observables,
we find {\em unsystematic}\/ discrepancies between the  
estimates given by the two methods:
for $m=100$ the discrepancies are of order 10\%,
and for $m=1000$ they are of order 2\%.
In other words, the size of these discrepancies is
roughly of order $\sim 1/\sqrt{m}$, and the sign is random;
this is exactly what one expects on theoretical grounds
for the statistical fluctuations in the estimators
\reff{O_error_bunch} and \reff{tau_int_error_bunch}.

%
% 2nd category 
% 
For the non-linear static observables, the agreement between the mean values 
coming from the standard and the bunch methods is exact
only in the case of the specific heat;
this is because we have used the same estimator  
$\overline{\cal E}$ in both methods.
For the correlation length, the mean values show small {\em systematic}\/
discrepancies between the two methods, of order 0.05--0.1 standard deviations
when $m=100$; the bunch method always gives a larger 
estimate than the standard method. In absolute value, these discrepancies 
range from $6 \times 10^{-4}$ (for $L=4$) to $2 \times 10^{-2}$ (for $L=256$). 
The same qualitative behavior is found when we repeat the analysis with 
$m=1000$ bunches. 
Now the differences in the mean value of the correlation length are 0.6--1.6 
standard deviations. In absolute value, they range from $8\times 10^{-3}$ 
(for $L=4$) to $4 \times 10^{-2}$ (for $L=64$), i.e.\ they are roughly 
one order of magnitude larger than in the $m=100$ case. 
Thus, the discrepancies in the 
mean value of the correlation length are systematic with a size of order 
$\sim m$;  this is exactly what one expects for the mean value
of a non-linear observable, which is afflicted by a bias of order $1/n$
in the standard method versus $1/\ell = m/n$ in the bunch method.
Regarding the error bars, we find {\em unsystematic}\/ discrepancies of order
10\% when $m=100$, and of order 2\% when $m=1000$.
They behave as the error bars of the static linear observables.

%
% 3rd category 
%
For the autocorrelation-time estimators,
we find {\em systematic}\/ differences between the two methods:
the estimates coming from the 
standard method are consistently smaller than those coming from the 
bunch method. For $m=100$, these differences are rather small compared 
to the statistical error bars ($\ltapprox 0.5$ standard deviations). 
When $m=1000$ these discrepancies
are much more relevant: their size is roughly 
two standard deviations. In absolute value, the discrepancies 
when $m=1000$ are one order of magnitude larger than when $m=100$.
Thus, we find a systematic bias of size $\sim m$,
again as expected for a quantity which exhibits a bias of order
$1/n$ versus $1/\ell$.
The error bars for the autocorrelation times are consistently 
larger for the standard method than for the bunch method,
except for the specific heat where the behavior is consistently 
the opposite one.
For $m=100$ bunches, these 
discrepancies are of order 15\%; and for $m=1000$ the size 
remains at the same level. Thus, the discrepancies for the error 
bars of dynamic observables do not depend much on $m$; rather they 
are of order $\sim 1$.

{}From the above discussion, we conclude that for the linear static observables 
the two methods show excellent agreement, both for the mean values
(trivial equality) and for the error bars (random discrepancies of
order $1/\sqrt{m}$).  The same holds for the error bars of the
non-linear static observables.
This confirms that the standard method is giving
accurate estimates of the error bars,
at least when the run length $n$ is large enough to provide 
a good determination of the autocorrelation time (which largely determines the
static-quantity error bar). 
This $1/\sqrt{m}$ dependence also confirms our theoretical prediction
that the standard method gives a {\em more accurate}\/ estimate of
the error bars than the bunch method.
Indeed, the bunch method can be considered roughly equivalent
to employing the standard method
{\em with the window width $M$ taken of order the bunch length $\ell$}\/;
but if $\ell \gg \tau_{{\rm int},{\cal O}}$
(as it {\em must}\/ be in the bunch method), this is an
{\em unnecessarily large}\/ window width, and thus leads to
unnecessarily large statistical fluctuations.

For the mean values of the non-linear observables (both static and dynamic),
we likewise confirm the validity of the standard method.
Once again the standard method is more reliable than the bunch method,
but for a different reason:  the bias of order $1/n$ is much smaller
than the bias of order $1/\ell = m/n$.
The latter bias becomes particularly serious when the number $m$
of bunches is large (as it {\em must}\/ be in order to get good
estimates of the {\em error bars}\/!).

Finally, the least understood piece is the determination of 
the error bar of the autocorrelation time:
our data from the bunch method suggest that the standard method
may be making a systematic error of order $15\%$.
Perhaps this systematic error (if indeed it is real)
arises from our neglect of the
contributions of the fourth-order cumulant $\kappa$
to the covariance \reff{cov_rho}.
This point definitely merits further investigation.

In summary, we think that the bunch method provides a good confirmation
of the estimates given by the standard method.
We shall hereafter consider the 
values given by the standard method to be the definitive ones,
except for the ratio $\tau_{{\rm int},{\cal E}}/C_H$
where the standard method does not yield any correct error bar.
In this latter case we shall use the central value coming from
the standard method
(which in fact agrees with the bunch-method value
to within 0.1--1\%; the discrepancies are of order 0.2 standard deviations), 
and the error bars coming from the bunch method
with 100 bunches (for $4 \leq L \leq 256$),
26 bunches (for $L=512$) or 55 bunches (for $L=1024$).
We shall also compute the upper bound on the error bar
coming from the standard method combined with the triangle inequality.
These results for $\tau_{{\rm int},{\cal E}}/C_H$ are shown in 
Table~\ref{table_tau_over_cv}.

% $\tau_{{\rm int},{\cal E}}/C_H$ we will use both the upper bound coming 
% from the triangle inequality (and the results of the standard method), 
% and the results with 100 bunches for $4 \leq L \leq 256$. For $L=512$ we
% could only use 26 bunches of length $\sim 10^4 \tau_{{\rm int},{\cal E}}$.  
% In this case the mean values coincide with those obtained 
% by the standard method, 
% and the discrepancies on the error bars vary from 20\% (linear observables) 
% to 10\% (non-linear observables). 
% For $L=1024$ we used 
% 55 bunches of length $\sim 10^3 \tau_{{\rm int},{\cal E}}$. Here, we also 
% checked that the mean values coming from both methods agree well within 
% one standard deviation, and the discrepancies on the error bars oscillate 
% between 2\% (linear observables) and 30\% (non-linear observables). It is 
% remarkable that in this case the second method with   
% a bunch length of $\ell \sim 10^3 \tau_{{\rm int},{\cal E}}$ 
% gives results in agreement with those coming from the standard method 
% (contrary to what happened for $4 \leq L \leq 64$). This is likely due to the 
% fact that here the total statistics is 20 times smaller than for 
% $4 \leq L \leq 64$, and the (larger) error bars are surely larger than the 
% bias in the mean values (which is of the same order as in those smaller 
% lattices with the same value of $\ell$).   
% The ratio $\tau_{{\rm int},{\cal E}}/C_H$ is shown in 
% Table~\ref{table_tau_over_cv} together with the error bar coming from the 
% bunch method and the bound coming from the standard method.  

\section{Data analysis} \label{sec_res}

For each quantity ${\cal O}$, we carry out a fit to the power-law
Ansatz ${\cal O} = A L^p$
using the standard weighted least-squares method.
As a precaution against corrections to scaling,
we impose a lower cutoff $L \ge L_{min}$
on the data points admitted in the fit,
and we study systematically the effects of varying $L_{min}$ on the 
estimates $A$ and $p$ and on the $\chi^2$ value.
In general, our preferred fit corresponds to the smallest $L_{min}$
for which the goodness of fit is reasonable
(e.g., the confidence level\footnote{
   ``Confidence level'' is the probability that $\chi^2$ would
   exceed the observed value, assuming that the underlying statistical
   model is correct.  An unusually low confidence level
   (e.g., less than 5\%) thus suggests that the underlying statistical model
   is {\em incorrect}\/ --- the most likely cause of which would be
   corrections to scaling.
}
is $\gtapprox$ 10--20\%),
and for which subsequent increases in $L_{min}$ do not cause the
$\chi^2$ to drop vastly more than one unit per degree of freedom. 

Our final estimates for static and dynamic critical exponents
are collected in Table \ref{table_results}.

\subsection{Static quantities} \label{sec_res_static}

There are a few exactly known results concerning the 2D 3-state Potts 
model. We know all the critical exponents \cite{Baxter,Alexander} 
and, in particular, the ratios  
\begin{eqnarray} 
{\gamma \over \nu}  &=& {26 \over 15} \approx 1.73333  \\[2mm]
{\alpha \over \nu}  &=& {2  \over 5}  = 0.4
\end{eqnarray} 
which are the relevant quantities we can directly estimate from our 
Monte Carlo data. 
The leading correction-to-scaling 
exponent is also known \cite{Nienhuis_82,Dotsenko}: 
\be 
\theta \;=\; {2 \over 3}  \;,
\ee
using the notation in which the correction term is
$|\beta-\beta_c|^\theta$ or $L^{-\theta/\nu}$.

We can write the 
singular part of the free energy as a function of the thermal field 
$t = \beta - \beta_c$, the ordering field $h$,
the leading irrelevant field $u$, and   
the linear size of the system $L$ as \cite{Privman}  
\be 
f_s(t,h,L) \;=\; L^{-d} F\left( t L^{1/\nu}, hL^{d-\beta/\nu}, 
                            uL^{-\theta/\nu} \right) \;,  
\label{fss_free_energy} 
\ee 
where $d$ is the dimensionality of the lattice. 
If we differentiate \reff{fss_free_energy} twice with respect to the thermal
field $t$ and then take the limit $t=h=0$, we get the specific heat at 
criticality on a finite lattice:  
\begin{subeqnarray}  
C_H(0,0,L) \sim {\partial^2 f_s \over \partial t^2}(0,0,L) &=& 
               L^{\alpha/\nu} G\left(0,0,uL^{-\theta/\nu} \right) \\ 
     &=& L^{\alpha/\nu} \left[ A + B L^{-\theta/\nu} + \cdots \right] \;,  
\end{subeqnarray} 
where $\alpha/\nu = 2/\nu - d$, 
$G$ is the second derivative of $F$ with respect to its first argument,
and the dots indicate subdominant corrections.   
Thus, the corrections to the specific heat at criticality are given by 
$L^{-\Delta}$ with $\Delta=\theta/\nu=4/5 = 0.8$. 
A similar analysis can be carried out for the magnetization and the 
susceptibility, giving again corrections proportional to $L^{-\Delta}$.

The energy $E$ is obtained by differentiating of the full free energy 
$f = f_s + f_{ns}$ with 
respect to the thermal field $t$. 
The contribution of the non-singular piece is believed to be trivial:
there is numerical evidence 
that $f_{ns}(t,L) = f_{ns}(t,\infty)$ \cite{Privman}.
In other words, this contribution has no $L$-dependence,
and gives merely the infinite-volume value of the energy
at the given temperature, $E(\beta,\infty)$. 
The contribution of the singular piece can be obtained by differentiating
\reff{fss_free_energy} once with respect to $t$;
it goes to zero like $L^{-d+1/\nu}$, which thus gives
the leading correction to scaling for the energy.
This correction is of order $ L^{-4/5}$,
which (by pure coincidence as far as we can tell)
is exactly the same order as the     
correction $L^{-\Delta}$ for the divergent static observables.  
Finally, we note that the energy of the 2D $q=3$ Potts model at criticality
is also exactly known 
\cite{Baxter} to be $E(\beta_c,\infty) = 1 + 1/\sqrt{3} \approx 1.577350$.

We can check these predictions by performing the fit 
$E - E(\beta_c,\infty) = A L^{-w}$. The quality of the fit is
very good already for $L_{min}=16$:
\be 
w = 0.803 \pm 0.002 
\ee
with $\chi^2=3.23$ (5 DF, level = 66\%). The agreement with 
the predicted exponent $2-1/\nu = 4/5 = 0.8$ is truly spectacular.
(Indeed, it is probably a coincidence that the agreement is {\em so}\/ good.
 In general, one can't expect to obtain anywhere near this accuracy\
 for correction-to-scaling exponents.)

The fits of the susceptibility to a power law $A L^{\gamma/\nu}$ are 
quite stable. Our preferred fit corresponds to $L_{min}=64$: 
\be 
  {\gamma \over \nu} \;=\; 1.73444 \pm 0.00043 \
\label{res_gamma_1024} 
\ee
with $\chi^2=4.09$ (3 DF, confidence level = 25\%). This result is 2.6 
standard deviations away from the exact value $\gamma/\nu = 1.73333$.  
This discrepancy could be due to corrections to scaling. We can try  
to fit our data to $A L^{26/15} (1 + BL^{-\Delta})$ with various choices for 
$\Delta=1.1,1.0,\ldots,0.1$ as well as   
$0\times \log$ (i.e., a correction $1/\log L$). We find that the best 
fits correspond to $\Delta \approx 0.8$, in agreement with the
theoretical prediction;  for $\Delta = 0.8$ and $L_{min}=8$ we 
obtain $\chi^2=6.55$ (6 DF, level = 36\%).
Surprisingly, in all these fits we 
find that the $\chi^2$ remains almost constant when $L_{min}$ is 
increased beyond our ``preferred'' value (and the confidence level 
consequently deteriorates): for example, with $L_{min}=256$ we get    
$\chi^2=3.37$ (1 DF, level = 7\%)  
for the pure power-law behavior and $\chi^2=3.91$ (1 DF, level = 5\%) 
for the fit with $\gamma/\nu=26/15$  
and $\Delta=0.8$. This suggests that the point with $L=1024$ 
is off by about two standard deviations (possibly because the error bar is 
underestimated\footnote{
   This is quite possible: though the total run length at $L=1024$
   is $7 \times 10^4 \tau_{{\rm int},{\cal E}}$,
   the {\em individual}\/ runs (on which the time-series analysis was
   performed) ranged in length from only $10^4 \tau_{{\rm int},{\cal E}}$
   to $2.5 \times 10^4 \tau_{{\rm int},{\cal E}}$;
   and with such short 
   runs the time-series analysis may not be completely reliable.
}). 
As a matter of fact, if we drop this point, we obtain 
a good power-law fit for $L_{min}=128$:  
\be
  {\gamma \over \nu} \;=\; 1.73337 \pm 0.00080 \
\label{res_gamma_512}
\ee
with $\chi^2=0.39$ (1 DF, level = 53\%). This result agrees excellently
with the exact value. If we impose the right power $\gamma/\nu=26/15$  
and try to fit the first correction-to-scaling exponent $\Delta$, we again find  
that the best fits correspond to $\Delta \approx 0.8$. For $L_{min}=8$ we 
get $\chi^2=3.84$ (5 DF, level = 57\%).

For the specific heat we find that the fits to power law $A L^{\alpha/\nu}$ 
are not stable at all: the confidence levels are horrible, and there  
is a clear trend towards smaller values of $\alpha/\nu$ as $L_{min}$ is 
increased. The least bad fit is obtained for $L_{min}=256$: 
\be 
  {\alpha \over \nu} \;=\; 0.4240 \pm 0.0030 
\label{res_alpha_1024} 
\ee 
with $\chi^2=3.80$ (1 DF, level = 5\%). This value is eight standard 
deviations away from the exact result $\alpha/\nu = 2/5 = 0.4$. 
Unlike the 4-state Potts model \cite{Nauenberg_Scalapino,Cardy80}, 
we do not expect 
multiplicative corrections to the leading term of the specific heat.  
We do, however, expect additive corrections to scaling of the form 
$A L^{2/5}(1 + BL^{-\Delta})$. If we try the same exponents $\Delta$ as 
in the susceptibility, we find a decent fit for $\Delta \approx 0.6$:
for $L_{min}=128$ we get $\chi^2=1.21$ (2 DF, level = 55\%). 
This value of the exponent $\Delta$ is not far from the expected value 0.8, 
but the cause of the discrepancy is unknown;
perhaps there is a large next-to-leading correction to scaling.

Finally, the second-moment correlation length $\xi$ is expected to behave 
linearly in $L$ as $L\rightarrow\infty$. In particular, the ratio 
$x\equiv\xi/L$ should approach a constant $x^\star$.  
We have tested this behavior. Already for $L_{min}=16$ our data are 
consistent with a constant value   
\be 
x^\star \;=\; \lim_{L\rightarrow\infty} {\xi(L) \over L}
       \;=\;  0.93235 \pm 0.00033
\label{res_x_1024} 
\ee 
with $\chi^2=8.08$ (6 DF, level = 23\%).
The fact that a good fit can be obtained 
with such a small $L_{min}$ implies 
that the corrections to scaling are very small for this observable.  
[If we fit to $\xi(L)/L = x^\star + B L^{-\Delta}$
 with $0.1 \ltapprox \Delta \ltapprox 1.5$,
 we find a very slight improvement in the goodness of fit,
 and the estimated $x^\star$ decreases somewhat
 (by less than 0.0008 if $\Delta \gtapprox 0.6$).
 But the estimated coefficient $B$ is consistent with zero
 at the $1.5\sigma$ level.]
As in the case of the susceptibility, we also find that the goodness of fit 
deteriorates as $L_{min}$ is increased (e.g., for $L_{min}=512$ we have 
$\chi^2=4.80$, 1 DF, level = 3\%)  
This might be due to the fact that the value for $L=1024$ is a little bit off 
(or its error bar is underestimated).   
If we drop this point, we obtain a good fit again for the same $L_{min}=16$:  
\be
x^\star  \;=\;  0.93229 \pm 0.00034
\label{res_x_512}
\ee
with $\chi^2=5.82$ (5 DF, level = 32\%). But now the fit with $L_{min}=256$ 
is reasonable ($\chi^2=1.24$, 1 DF, level = 27\%).  
We remark that the value $x^\star$ should in principle be calculable by 
conformal field theory; we hope that someone will perform this 
calculation and test our results \reff{res_x_1024}/\reff{res_x_512}.

\subsection{Dynamic quantities} \label{sec_res_dynamic}

In this section we are going to fit the autocorrelation times for the 
observables ${\cal O} = {\cal E}$, ${\cal E}'$, ${\cal N}$ and ${\cal M}^2$   
to a simple power law 
$\tau_{{\rm int},{\cal O}} = A L^{z_{{\rm int},{\cal O}}}$.

Let us start with the energy ${\cal E}$.  
The fit $\tau_{{\rm int},{\cal E}} = A L^{z_{{\rm int},{\cal E}}}$  
is not very stable: the estimate of 
the power decreases systematically as $L_{min}$ is increased,  
and the $\chi^2$ is poor until $L_{min}=128$ where the 
estimate stabilizes  
within errors (this behavior suggests that there are strong 
corrections to scaling). Our preferred fit corresponds to $L_{min}=128$: 
\be 
z_{{\rm int},{\cal E}} \;=\; 0.515 \pm 0.006 
\label{res_z_1024} 
\ee 
with $\chi^2=0.44$ (2 DF, level = 80\%). This value is greater than our 
estimate for $\alpha/\nu$; thus, the Li--Sokal 
bound \reff{Li_Sokal_bound} holds, 
though apparently not as a strict equality.
To compare our result with the one reported  
in \cite{Li_Sokal}, we redid the fit using only the data with $L\leq 256$.  
The fit again shows a systematic decrease of the exponent as $L_{min}$ 
is increased, as well as a poor $\chi^2$, until $L_{min}=64$ where we get 
$z_{{\rm int},{\cal E}} = 0.531 \pm 0.005$ with $\chi^2=2.67$ 
(1 DF, level = 10\%). This is consistent with the result 
$z_{{\rm int},{\cal E}} = 0.55 \pm 0.02$ reported in \cite{Li_Sokal}, 
but our error bar is one-fourth of theirs. The slightly higher 
estimate of $z_{{\rm int},{\cal E}}$ in \cite{Li_Sokal} seems to arise 
from corrections to scaling induced by their choice
$L_{min}=16$. Actually, the best fit to the data in \cite{Li_Sokal}
corresponds to $L_{min}=32$ and gives 
$z_{{\rm int},{\cal E}} = 0.52 \pm 0.02$ with $\chi^2=3.93$ 
(2 DF, level = 14\%).

The fit of the autocorrelation time for the bond occupation ${\cal N}$
follows the same pattern: the estimate of the power decreases strongly
as $L_{min}$ increases, and the $\chi^2$ is initially horrible; 
eventually the power stabilizes within errors,
and the $\chi^2$ becomes reasonable.
Our preferred fit is $L_{min}=128$:  
\be 
z_{{\rm int},{\cal N}} \;=\; 0.529 \pm 0.006
\ee
with $\chi^2=0.71$ (2 DF, level = 70\%). The difference with respect to
$z_{{\rm int},{\cal E}}$ is only 2.3 standard deviations,
consistent with the theoretical prediction \reff{equal3}
that $z_{{\rm int},{\cal N}} = z_{{\rm int},{\cal E}}$.
To check this result,
we studied the ratio 
$\tau_{{\rm int},{\cal N}}/\tau_{{\rm int},{\cal E}}$.
Since the standard time-series-analysis method would 
give only a upper bound on the right error bar for this quantity,
we used instead the error bar provided by the ``bunch'' method.  
Among all the Ans\"atze we tried, only one gave a good 
fit for $L_{min}=128$: asymptotically constant with corrections 
$\sim L^{-1/4}$ ($\chi^2 = 0.75$, 2 DF, level = 69\%).
Thus, we conclude that in the SW algorithm the 
dynamic critical behavior of the energy and the bond density are the same:
$z_{{\rm int},{\cal N}} = z_{{\rm int},{\cal E}}$.

In the same way, we considered the nearest-neighbor connectivity ${\cal E}'$. 
The pattern of the fit is the same 
as above, and our preferred fit again corresponds to $L_{min}=128$:  
\be
z_{{\rm int},{\cal E'}} \;=\; 0.514 \pm 0.006
\ee
with $\chi^2=0.47$ (2 DF, level = 79\%). This time, the agreement with 
$z_{{\rm int},{\cal E}}$ is extremely good,
confirming the theoretical prediction \reff{equal4}
that $z_{{\rm int},{\cal E'}} = z_{{\rm int},{\cal E}}$.

Finally, a similar behavior is observed in the  
fit for the autocorrelation time of the squared magnetization 
${\cal M}^2$:
the estimate of the power shows a
clear trend towards smaller values as $L_{min}$ is increased,
and the $\chi^2$ is initially poor.
Our preferred fit is $L_{min} = 128$, for which we get 
\be 
z_{{\rm int},{\cal M}^2} \;=\; 0.475 \pm 0.006
\ee
with $\chi^2=0.36$ (2 DF, level = 84\%). 
This power is 
slightly smaller than $z_{{\rm int},{\cal E}}$ (the difference is seven 
standard deviations). It would be interesting to know whether this difference 
is significant or not. Let us consider the ratio 
$\tau_{{\rm int},{\cal M}^2}/\tau_{{\rm int},{\cal E}}$,
again using the error bar provided by the ``bunch'' method.
We tried to fit this ratio to various Ans\"atze, 
but only two gave good results for $L_{min}=32$: a pure power-law 
behavior $A L^p$ with $p=-0.0409\pm0.0007$
($\chi^2=0.57$, 5 DF, level = 97\%);  
and a constant plus corrections of the type $\sim L^{-1/16}$
($\chi^2=0.41$, 5 DF, level = 98\%). 
We are therefore unable to resolve whether
$z_{{\rm int},{\cal M}^2}$ is exactly equal to $z_{{\rm int},{\cal E}}$
or not;  but if it is not equal, then it is only very slightly smaller
($z_{{\rm int},{\cal M}^2} - z_{{\rm int},{\cal E}} \approx -0.04$).

In conclusion, we have shown numerically
that the energy ${\cal E}$, the bond occupation ${\cal N}$,
and the nearest-neighbor connectivity ${\cal E'}$
all have the same dynamic critical exponent $z_{\rm int}$,
as was ``almost proved'' in Section~\ref{sec_setup}.
On the other hand, the observable ${\cal M}^2$ has 
a {\em similar but perhaps not identical}\/ dynamic critical behavior:
$z_{{\rm int},{\cal M}^2}$ may coincide with $z_{{\rm int},{\cal E}}$,
or it may be slightly smaller.

\subsection{Analysis of the Li--Sokal bound} \label{sec_res_li_sokal}

In the previous subsection we have seen that all the observables considered 
have the same (or, in the case of ${\cal M}^2$, almost the same) 
dynamic critical exponent, and the common $z_{\rm int}$ is strictly 
larger than $\alpha/\nu$. This implies that the Li--Sokal bound 
\reff{Li_Sokal_bound} is {\em not sharp}\/. However, the difference 
$z_{\rm int} - \alpha/\nu \approx 0.115$ is actually not very large. 
There are a few arguments in favor of a more detailed analysis: 

\begin{enumerate} 

\item[i)] 
The power-law fit to the specific heat was not very good: the estimated value 
of $\alpha/\nu$ decreased as $L_{min}$ increased, and we were unable to reach 
(within statistical errors) the exact value $\alpha/\nu = 2/5$. If we compare 
the {\em observed}\/  values of $z_{\rm int}$  and $\alpha/\nu$, we find that
$z_{\rm int} - \alpha/\nu$ is only $\approx 0.09$.   

\item[ii)] 
The power-law fits to the autocorrelation times also exhibited 
this trend to smaller
values as $L_{min}$ increased. Even though the fits seemed stable for 
$L_{min}=128$, this might well be an artifact due to the large error bars  
associated with the largest lattices ($L\geq512$) compared to the smallest 
ones ($L\leq256$).  

\item[iii)] 
In \cite{Salas_Sokal_LAT95,Salas_Sokal_AT} it was shown that differences 
$z_{\rm int} - \alpha/\nu$ of order 0.1 could actually 
be due to multiplicative logarithmic corrections.  
Indeed, such a multiplicative logarithmic correction was found to be a 
likely scenario even for models {\em not}\/ having such a 
logarithmic correction in the specific heat. 

\end{enumerate}

Here, we will follow the approach of 
Refs.~\cite{Salas_Sokal_LAT95,Salas_Sokal_AT}, and consider the ratio  
$\tau_{{\rm int},{\cal E}}/C_H$; we shall use the error bars 
obtained from the ``bunch'' method 
(see Table~\ref{table_tau_over_cv}).\footnote{   
    In this section we only consider the energy ${\cal E}$,
    as the other observables 
    have the same (or very slightly smaller) dynamic critical exponent. 
}  
We have tried to fit this ratio to different Ans\"atze: 

\begin{itemize} 

\item[1)] 
A pure power-law 
behavior $A L^p$. 

\item[2)] 
A logarithmic growth, either as $A + B \log L$ or as 
$A \log^p L$. 

\item[3)] 
Asymptotically constant with additive corrections to scaling 
$A + B L^{-\Delta}$. We have considered the cases 
$\Delta=2,1,{1\over 2},{1\over 4},{1 \over 8}$, and  
$0 \times \log$ (i.e., $\tau_{{\rm int},{\cal E}}/C_H = A + B/\log L$).  

\end{itemize} 

Among all these Ans\"atze, only two were reasonably good.  
The best one is the simple power-law behavior $A L^p$. For 
$L_{min}=32$ we obtain 
\be 
p\;=\;0.084\pm0.002 
\ee
with $\chi^2=1.72$ 
(4 DF, level = 79\%). When $L_{min}>32$, the value of the power $p$ stays 
stable, and the $\chi^2$ decreases slowly and consistently.

The second-best fit corresponds to the logarithmic growth $A + B\log L$. 
For $L_{min}=64$, we get $\chi^2=1.93$ (3 DF, level = 59\%). Again, the 
estimates are stable for $L_{min}>64$ and the $\chi^2$ is reasonable.  
However, this Ansatz seems to be slightly inferior to the power-law 
fit: both $L_{min}$
and the $\chi^2$ are greater than in the power-law case. 
To test the logarithmic Ansatz, we can impose the known critical behavior
of the specific heat and perform the fit 
$\tau_{{\rm int},{\cal E}} = L^{2/5}(A + B \log L)$. A reasonably good 
result is obtained for $L_{min}=32$, giving $\chi^2=1.54$ (4 DF, level = 82\%).

The asymptotically constant fits are always horrible, unless one take 
$L_{min}=256$ or larger. The only semi-exception is 
$\Delta={1\over 8}$, which gives a tolerable fit 
($\chi^2=5.54$, 3 DF, level = 14\%) already for $L_{min}=64$.

Of course 
for $L_{min}=256$ we obtain reasonably good fits for all  
Ans\"atze. But this is because the error bars for $L>256$ are so large
that we are unable to distinguish between the various scenarios.

In summary, we have shown that there are only two likely scenarios for the 
ratio $\tau_{{\rm int},{\cal E}}/C_H$: 
\be 
\tau_{{\rm int},{\cal E}}/C_H  \;\sim\; \left\{ \begin{array}{l}  
           A L^p \quad  \hbox{with $p =  0.084 \pm 0.002$} \\ 
           A + B \log L           \\ 
           \end{array} \right. 
\ee
or equivalently
\be
\tau_{{\rm int},{\cal E}} \;\sim\; \left\{ \begin{array}{l}
           A L^{z_{{\rm int},{\cal E}}} \quad  
             \hbox{with $z_{{\rm int},{\cal E}}=0.515\pm0.006$} \\
           L^{2/5}(A + B \log L)       \\
           \end{array} \right.
\ee 
The first scenario implies that the Li--Sokal bound \reff{Li_Sokal_bound} is 
a strict inequality, with
$p \equiv z_{{\rm int}, {\cal E}} - \alpha/\nu \approx 0.08$--0.12;
while the second scenario means that this bound is violated only 
by a logarithm (i.e., the bound is sharp modulo logarithms). 
Thus, the same ``dynamic continuity'' we found interpolating between the 
2-state and the 4-state Potts models along the AT self-dual line is also 
found along the $q$-state Potts model critical line. Moreover, the 
power $p$ found is midway between the power found for the Ising model 
($p \approx 0.05$) and the one found for the 4-state Potts model
($p \approx 0.12$) \cite{Salas_Sokal_AT}.

\subsection{Further discussion of the Li--Sokal bound} 
\label{sec_res_lisokal_bis}

The proof of the Li--Sokal bound (Section~\ref{sec_setup_lisokal}) is 
based on computing the autocorrelation function for the observable 
${\cal N}$ at time lags 0 and 1 in terms of static observables, and 
then exploiting the inequality 
$\rho_{\cal NN}(t) \geq \rho_{\cal NN}(1)^{|t|}$.  
The apparent non-sharpness of this bound indicates that the large-$t$ 
behavior of $\rho_{\cal NN}(t)$ is {\em not}\/ fully predicted by its 
behavior at $t=1$. Otherwise put, 
if we define the {\em initial}\/ autocorrelation time 
\be 
\tau_{{\rm init},{\cal O}} \;\equiv\; {1\over2} \,\times\,
 {1 + \rho_{\cal OO}(1) \over 1 - \rho_{\cal OO}(1) } \;,  
\label{def_tau_init} 
\ee
then the Li--Sokal bound 
\be 
\tau_{{\rm int},{\cal N}} \;\geq\; \tau_{{\rm init},{\cal N}} = 
{p \over 1-p} {C_H \over E} + {1 \over 2} 
\label{equal_tauinit} 
\ee
is apparently {\em not sharp}\/: that is, 
$\tau_{{\rm int},{\cal N}}$ and $\tau_{{\rm init},{\cal N}}$ diverge with 
{\em different}\/ critical exponents 
$z_{{\rm int},{\cal N}}$ and $z_{{\rm init},{\cal N}}= \alpha/\nu$.

One might ask whether the situation could be improved by computing 
$C_{\cal NN}(t)$ exactly at (for example) time lag $t=2$ or $t=3$. 
Because of the identities \reff{check_c_1}/\reff{check_c_2}, 
this would be equivalent to carrying out the Li--Sokal proof using 
${\cal E}$ or ${\cal E}'$ as the test observable in place of ${\cal N}$. 
At present we do not know how carry out this computation --- the 
trouble is that we do not know how to express 
$\<\gamma_{ij}\gamma_{kl}\>$ in terms of spin observables --- but we can 
nevertheless test numerically whether we {\em would}\/ obtain a better bound 
on $z_{{\rm int},{\cal E}}$ and/or $z_{{\rm int},{\cal E}'}$ if we 
{\em could}\/ carry out this computation.

We thus computed $\tau_{{\rm init},{\cal O}}$ for 
${\cal O} = {\cal N}$, ${\cal E}$, ${\cal E}'$, with error bars 
given by \reff{cov_rho}\footnote{ 
   Since the autocorrelation function for these observables is almost 
   exactly a pure exponential, it suffices to perform the sums in 
   \protect\reff{cov_rho} {\em analytically}\/ and then insert the 
   appropriate value of $\tau$ (see \cite{Salas_Sokal_AT}). 
};
we then tried various fits, among others the fit 
$\tau_{{\rm init},{\cal O}}/C_H = A L^r$.\footnote{ 
   For the error bars on $\tau_{{\rm init},{\cal O}}/C_H$, we used for 
   simplicity the triangle inequality rather than the (more correct) 
   bunch method. 
}

For ${\cal O}={\cal N}$ we have the identity 
\begin{subeqnarray} 
{ \tau_{{\rm init},{\cal N}} \over C_H } &=& 
{p \over 1-p} \; {1 \over E} + {1 \over 2 C_H} \\ 
 &=& \left[ {p \over 1-p} {1 \over E(\beta,\infty)} + a L^{-4/5} 
              + \cdots \right] 
   + \left[ b L^{-2/5} + cL^{-6/5} + \cdots \right] \;,  
\end{subeqnarray} 
where $E(\beta,\infty)$ is the infinite-volume energy at inverse
temperature $\beta$; and
we have taken into account the leading terms and the corrections to scaling 
discussed in Section~\ref{sec_res_static}. 
At criticality $\beta=\beta_c = \log(1+\sqrt{3})$
[hence $p=p_c = \sqrt{3}/(1 + \sqrt{3})$]
we know \cite{Baxter} the exact constant term:
\be 
{ p_c \over 1-p_c} \, {1 \over E(\beta_c,\infty)} \;=\; {3 \over 1 + \sqrt{3}} 
\;\approx\; 1.098076 \;. 
\ee 
Indeed, if we 
fit our data to the Ansatz 
$\tau_{{\rm init},{\cal N}}/C_H - 3/(1 + \sqrt{3}) = b L^{-s}$ we 
obtain a good fit for $L_{min}=16$, and the estimate is 
\be 
s \;=\; 0.415 \pm 0.017 
\ee
($\chi^2 = 1.92$, 3 DF, level = 86\%), in excellent agreement with the 
theoretical prediction. Of course, we could try also the more primitive 
fit 
$\tau_{{\rm init},{\cal N}}/C_H = A L^r$, and we get a good fit for 
$L_{min}=128$: 
\be 
r \;=\; -0.010 \pm 0.002 
\ee
($\chi^2=0.90$, 2 DF, level = 64\%), which is very close to the exact 
result $r=0$. 
All this is trivially to be expected, since our raw data for
$\rho_{\cal NN}(1)$, $C_H$ and $E$ are in good agreement with the
theoretical identity \reff{check_rho}.

The analysis becomes non-trivial when we look at ${\cal O}={\cal E}$ 
and ${\cal E'}$.
For ${\cal O}={\cal E}$ we fit 
$\tau_{{\rm init},{\cal E}}/C_H = A L^r$, and get a good fit for 
$L_{min}=64$: 
\be 
r \;=\; 0.003 \pm 0.003 
\ee
($\chi^2=0.91$, 3 DF, level = 82\%). For ${\cal O}={\cal E}'$ 
the analogous fit is good for $L_{min}=32$: 
\be 
r \;=\; 0.016 \pm 0.001 
\ee
($\chi^2=0.51$, 4 DF, level = 97\%). These exponents $r$ are a factor of 
$\approx 5$ smaller than the exponent estimates for 
$p \equiv z - \alpha/\nu$ found in Section~\ref{sec_res_li_sokal}. 
This suggests that 
$\tau_{{\rm init},{\cal E}}/C_H$ and $\tau_{{\rm init},{\cal E}'}/C_H$ in 
fact tend to {\em finite constants}\/ as $L\rightarrow\infty$ 
(as we know rigorously to be the case for 
$\tau_{{\rm init},{\cal N}}/C_H$), and that the apparent non-sharpness 
of the Li--Sokal bound cannot be remedied by using 
${\cal E}$ or ${\cal E}'$ as a test observable in place of ${\cal N}$. 
Rather, the non-sharpness of the bound arises from the fact that the 
long-time behavior of $\rho_{\cal NN}(t)$ is not sufficiently well
predicted by its behavior at {\em any}\/ small time.

\subsection{Exponential autocorrelation time} \label{sec_res_tauexp}

The exponential autocorrelation time for an observable ${\cal O}$
is defined as\footnote{
   For a general Markov chain,
   the ``$\lim$'' should strictly speaking be replaced by ``$\limsup$'',
   and $\rho_{\cal O O} (t)$ should be replaced by its absolute value.
   But here it can be {\em proven}\/ that the limit really exists,
   and that $\rho_{\cal O O} (t) \ge 0$ for all $t$;
   this follows from the spectral representation \reff{spectral_rep}
   [or rather its analogue for ${\cal O}$].
}
\be
\tau_{{\rm exp},{\cal O}}  \;=\;
   \lim_{t \to \infty}  {-|t| \over  \log  \rho_{\cal O O} (t)}
   \;.
\ee
This autocorrelation time measures the decay rate of the
``slowest mode'' of the system,
provided that this mode is not orthogonal to $\cal O$.

The critical behavior of
$\tau_{{\rm exp},{\cal O}}$ is, in general, different from the behavior of
$\tau_{{\rm int},{\cal O}}$.
This fact can be seen from the standard dynamic finite-size-scaling
Ansatz for the autocorrelation function $\rho_{\cal OO}(t)$:
\be
\rho_{\cal OO}(t;L) \;\approx\; |t|^{-p_{\cal O}} h_{\cal O}\!\left(
          {t \over \tau_{{\rm exp},{\cal O}} }; {\xi(L) \over L} \right) \; .
\label{fss_rho}
\ee
(Here the dependence on the coupling constants has been suppressed
for notational simplicity.)  Summing \reff{fss_rho} over $t$, it follows that
\be
\tau_{{\rm int},{\cal O}} \;\sim\;
 \tau_{{\rm exp},{\cal O}}^{1 - p_{\cal O}} \; ,
\ee
or equivalently,
\be
z_{{\rm int},{\cal O}}  \;=\;  (1 - p_{\cal O}) z_{{\rm exp},{\cal O}} \; .
\ee
Thus, only when $p_{\cal O}=0$ do we have
$z_{{\rm int},{\cal O}} = z_{{\rm exp},{\cal O}}$
\cite{Sokal_Lausanne}. 
In this latter case the Ansatz \reff{fss_rho} can be rewritten in
the equivalent form
\be
\rho_{\cal OO}(t;L) \;\approx\; \widehat{h}_{\cal O}\!\left(
          {t \over \tau_{{\rm int},{\cal O}} }; {\xi(L) \over L} \right) \; .
\label{fss_rho_p=0}
\ee

\begin{figure}
% \epsffile[100 166 565 690]{rho_N_error_all.ps}
\epsfxsize=400pt\epsffile{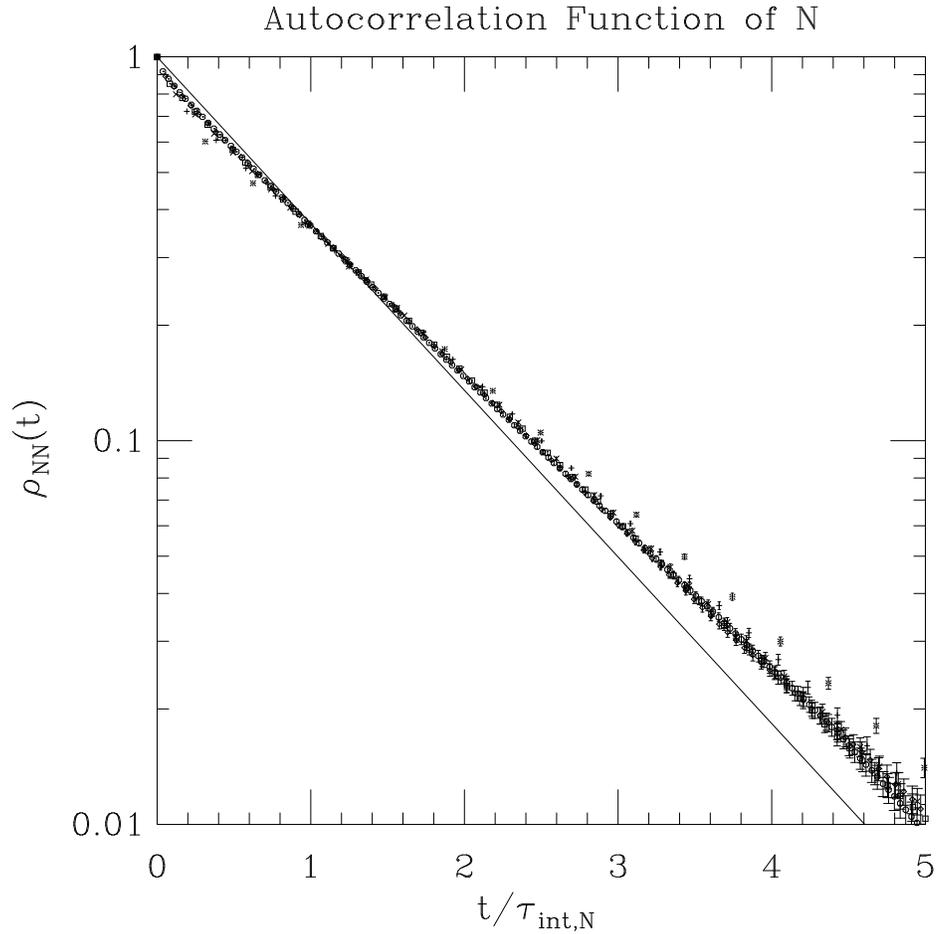}
  \caption{
  Plot of $\rho_{\cal NN}(t)$ versus $t/\tau_{{\rm int},{\cal N}}$ for 
  $4 \leq L \leq 128$. The different symbols denote the different 
  lattice sizes: $L=4$ ($\ast$), $L=8$ ($+$), $L=16$ ($\times$), 
  $L=32$ ($\Box$), $L=64$ ($\Diamond$), and $L=128$ ($\circ$). 
  We have also depicted the line corresponding to the pure 
  exponential $\rho_{\cal NN}(t) = \exp(-t/\tau_{{\rm int},{\cal N}})$.   
  }
\label{figure_fss_rhoN}
\end{figure}

\begin{figure}
% \epsffile[100 166 565 690]{rho_E_error_all.ps}
\epsfxsize=400pt\epsffile{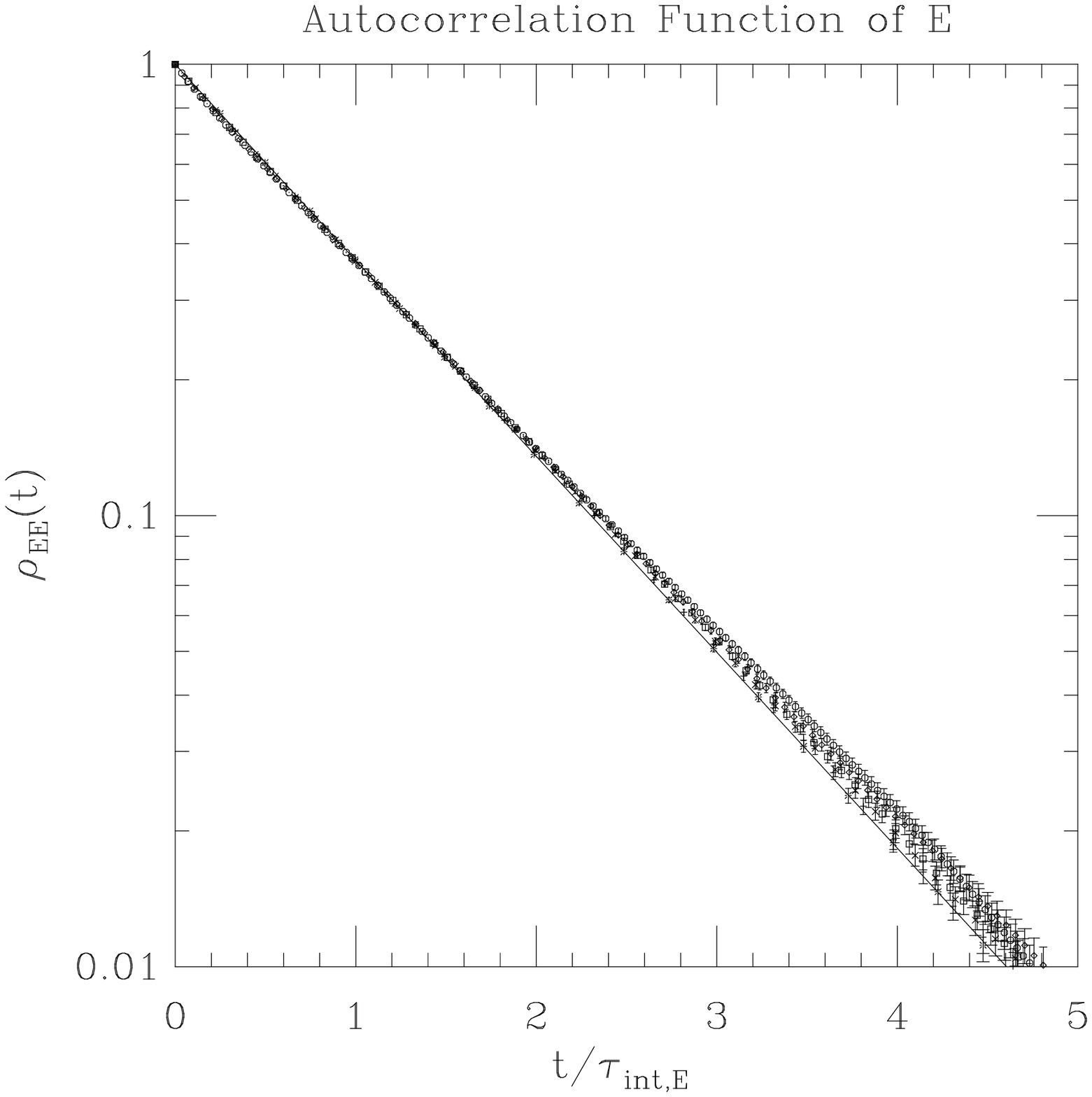}
  \caption{
  Plot of $\rho_{\cal EE}(t)$ versus $t/\tau_{{\rm int},{\cal E}}$ for
  $4 \leq L \leq 128$. The symbols are as in 
  Figure~\protect\ref{figure_fss_rhoN}.   
  We have also depicted the line corresponding to the pure  
  exponential $\rho_{\cal EE}(t) = \exp(-t/\tau_{{\rm int},{\cal E}})$.  
  }
\label{figure_fss_rhoE}
\end{figure}

\begin{figure}
% \epsffile[100 166 565 690]{rho_EP_error_all.ps}
\epsfxsize=400pt\epsffile{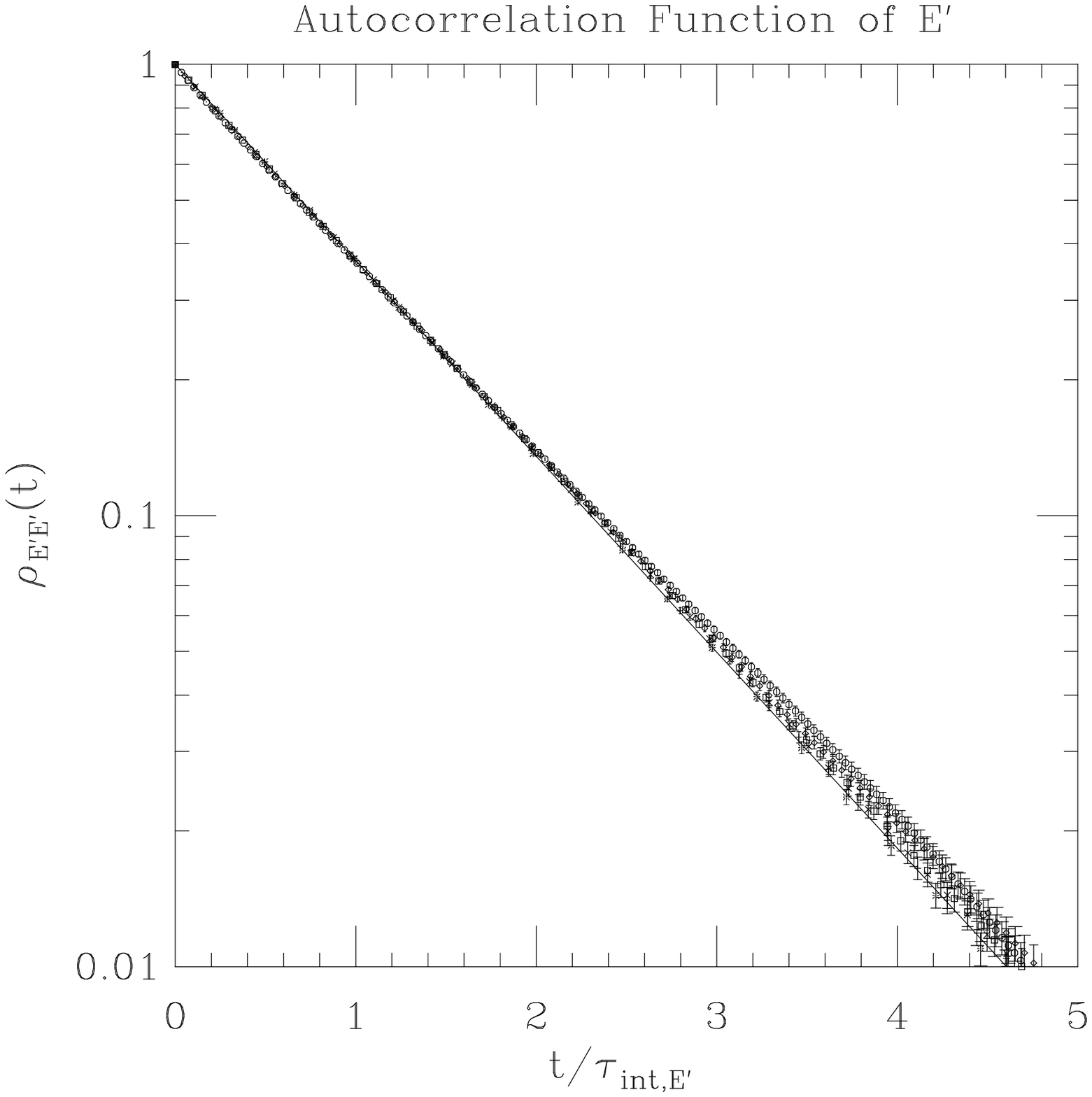}
  \caption{
  Plot of $\rho_{\cal E'E'}(t)$ versus $t/\tau_{{\rm int},{\cal E'}}$ for
  $4 \leq L \leq 128$. The symbols are as in  
  Figure~\protect\ref{figure_fss_rhoN}.  
  We have also depicted the line corresponding to the pure  
  exponential $\rho_{\cal E'E'}(t) = \exp(-t/\tau_{{\rm int},{\cal E'}})$.  
  }
\label{figure_fss_rhoEP}
\end{figure}

To test this latter Ansatz,
we have plotted $\log \rho_{\cal OO}(t)$ versus
$t/\tau_{{\rm int},{\cal O}}$ for  
the observables ${\cal O}=$ ${\cal N}$ (Figure~\ref{figure_fss_rhoN}), 
${\cal E}$ (Figure~\ref{figure_fss_rhoE}) and 
${\cal E}'$ (Figure~\ref{figure_fss_rhoEP}).
On each figure we have plotted 
the data coming from different lattice sizes ($4 \leq L \leq 128$)
with different symbols;
the error bars are computed from \reff{cov_rho},
using for simplicity the approximation \cite{Salas_Sokal_AT}
that the decay is a pure exponential (which is here almost exact).
On each graph, we have also depicted for reference
a line corresponding to the pure exponential 
$\rho_{\cal OO}(t) = e^{-t/\tau_{{\rm int},{\cal O}}}$.  
In these plots we have omitted the data for $L \ge 256$
for the sake of visual clarity;
these data agree well with the curve found for smaller $L$,
but with huge statistical error bars.

For ${\cal O}={\cal N}$ we see that the data coming from $16\leq L \leq 128$ 
collapse well onto a single curve. 
The lattices $L=4,8$ show slight systematic deviations
from this limiting curve:  these deviations are negative for
$t/\tau_{{\rm int},{\cal N}} \ltapprox 1.5$
and positive for
$t/\tau_{{\rm int},{\cal N}} \gtapprox 1.5$.
This trend continues for $16 \le L \le 128$,
but the deviations are in most cases smaller than the error bars 
(especially for the larger lattices).

A similar behavior is found for ${\cal O}={\cal E}$ 
(Figure~\ref{figure_fss_rhoE}) and ${\cal O}={\cal E}'$ 
(Figure~\ref{figure_fss_rhoEP}), but the corrections to scaling
are much weaker, and their sign is opposite from those seen for ${\cal N}$.

Thus, within statistical 
errors, we have found that the Ansatz \reff{fss_rho_p=0} is satisfied.
This implies that the
integrated and exponential autocorrelation times for these three
observables have exactly the same dynamic critical exponent,
i.e.\ $z_{{\rm int}, {\cal O}} = z_{{\rm exp}, {\cal O}}$
for ${\cal O} = {\cal N}, {\cal E}, {\cal E'}$.
This equality does {\em not}\/ hold as a general rule
in the theory of dynamic critical phenomena
\cite{Sokal_Lausanne,Sokal_LAT90},
but it does appear to hold for algorithms of Swendsen--Wang type.

Finally, we find that $\rho_{\cal NN}(t)$ differs slightly but noticeably
from the pure exponential $e^{-t/\tau_{{\rm int},{\cal N}}}$.
The discrepancy from a pure exponential becomes smaller
when we consider the other two observables ${\cal E}$ and ${\cal E}'$. 
This is to be expected theoretically from the relations
${\cal E} \sim P_{\rm bond} {\cal N}$ and
${\cal E'} \sim P_{\rm spin} {\cal E}$
[cf.\ \reff{eq2.14}/\reff{eq2.35}]:
each action of $P_{\rm bond}$ or $P_{\rm spin}$ helps to ``purify''
the slowest mode, so that the autocorrelation function
becomes closer to a pure exponential.
On the other hand, the identities \reff{inequal2}/\reff{inequal3}
imply\footnote{
   Using the fact that $\rho_{\cal NN}(1) \to 1$
   and $\tau_{{\rm int}, {\cal N}} \to \infty$
   as $L \to \infty$, as follows from the Li--Sokal identity
   \reff{check_rho} and the fact that the specific heat $C_H$
   is divergent.
}
that the limiting (scaling) functions
$\widehat{h}_{\cal N}$, $\widehat{h}_{\cal E}$ and $\widehat{h}_{\cal E'}$
are {\em identical}\/ (assuming they exist at all);
this means that at least some of the curves in
Figures~\ref{figure_fss_rhoN}--\ref{figure_fss_rhoEP}
have not yet reached their scaling limit.

Finally, a crude fit suggests that
$\tau_{{\rm int},{\cal E}}/\tau_{{\rm exp},{\cal E}} \approx 0.96$,
in agreement with the idea that $\rho_{\cal EE}(t)$ is almost but not quite
a pure exponential.

%                                                                               
%END TEXT SECTIONS
%
\section*{Acknowledgments}

We wish to give our {\em big}\/ thanks to Wolfhard Janke
for pointing out an error in an earlier version of this paper.
We also thank the Cornell Theory Center for the computer time needed to 
complete this project.
The authors' research was supported in part 
by U.S.\ National Science Foundation grants DMS-9200719
and PHY-9520978, and by Metacenter grant MCA94P032P.

\newpage
\renewcommand{\baselinestretch}{1}
\large\normalsize
%
%
%
%%%%%%%%%%%%   references  %%%%%%%%%%%%%%%%%%%%%%%%
%

\clearpage

%%%%%%%%%%%% START OF TABLES %%%%%%%%%%%%%
\newpage 
\def\kk{\phantom{1}} 

%% Table generated by sw_potts3_make_table_normal.pl
%% Data for 3-state Potts model @ Tc
%% TABLE 1: STATIC QUANTITIES 
%% 
\begin{table}%[th]  
%\centering 
%\vspace*{-0.5in} % Move table upwards
\addtolength{\tabcolsep}{-1.0mm}
\hspace*{-1.0cm}    % Move table leftwards
%\protect\footnotesize 
\begin{tabular}{|r|r|r@{$\,\pm\,$}r|r@{$\,\pm\,$}r|%
                     r@{$\,\pm\,$}r|r@{$\,\pm\,$}r|}% 
\hline\hline  \\[-0.5cm] 
\multicolumn{1}{|c|}{$L$}    & 
\multicolumn{1}{c|}{$MCS$}   & 
\multicolumn{2}{c|}{$\chi$}  & 
\multicolumn{2}{c|}{$C_H$}   & 
\multicolumn{2}{c|}{$\xi$}   & 
\multicolumn{2}{c|}{$E$}   \\[0.1cm] 
\hline\hline 
    4 & {\rm exact}
      &\multicolumn{2}{l|}{\kk\kk\kk\kk12.204711}
      &\multicolumn{2}{l|}{\kk1.496719}
      &\multicolumn{2}{l|}{\kk\kk3.862380} 
      &\multicolumn{2}{l|}{1.710062}  \\ 
\hline 
    4 & 4.9 &    12.2093 &   0.0054 &     1.4952 &   0.0021 &     3.8635 &   0.0043 &  1.71051 &  0.00039 \\ 
    8 & 6.9 &    41.3527 &   0.0204 &     2.4712 &   0.0031 &     7.5419 &   0.0072 &  1.65391 &  0.00026 \\ 
   16 & 9.9 &   138.4491 &   0.0723 &     3.7162 &   0.0046 &    14.9267 &   0.0134 &  1.62070 &  0.00016 \\ 
   32 & 14.9 &   462.7014 &   0.2408 &     5.2941 &   0.0064 &    29.8509 &   0.0250 &  1.60223 &  0.00010 \\ 
   64 & 19.9 &  1542.6921 &   0.8428 &     7.3839 &   0.0094 &    59.6964 &   0.0501 &  1.59163 &  0.00006 \\ 
  128 & 29.9 &  5135.9512 &   2.7481 &    10.1007 &   0.0127 &   119.3401 &   0.0952 &  1.58552 &  0.00003 \\ 
  256 & 40.8 & 17082.8221 &   9.2802 &    13.7116 &   0.0176 &   238.4565 &   0.1888 &  1.58202 &  0.00002 \\ 
  512 & 12.9 & 56760.2838 &  65.0770 &    18.4752 &   0.0507 &   475.9494 &   0.7779 &  1.58001 &  0.00003 \\ 
 1024 & 5.5 & 189676.5530 & 387.3500 &    24.5281 &   0.1247 &   958.8929 &   2.7875 &  1.57893 &  0.00003 \\ 
\hline 
$\infty$ & {\rm exact}
      &\multicolumn{2}{l|}{}
      &\multicolumn{2}{l|}{}
      &\multicolumn{2}{l|}{} 
      &\multicolumn{2}{l|}{1.577350}  \\ 
[0.1cm] 
\hline\hline
\end{tabular}
\caption{Static data from the MC simulations at the critical point  
of the 3-state Potts model.  
For each lattice size ($L$), we include the number of performed 
measurements ($MCS$) in units of $10^6$, the susceptibility ($\chi$), the 
specific heat ($C_H$), the second-moment correlation length ($\xi$), and 
the energy ($E$). The 
quoted errors correspond to one standard deviation (i.e. confidence level 
$\approx$ 68\%). The first row (``exact'') gives the {\em exact}\/ results 
for the $4\times4$ lattice, and the last row (``$L=\infty$'') gives the 
{\em exact}\/ energy in the limit $L\rightarrow\infty$.
}
\label{table_static_full}
\end{table}

%% Table generated by sw_potts3_make_table_normal.pl
%% Data for 3-state Potts model @ Tc
%% TABLE 2: DYNAMIC QUANTITIES 
%% 
\begin{table}%[hb]  
\centering 
%\vspace*{-0.5in} % Move table upwards
\addtolength{\tabcolsep}{-1.0mm}
%\hspace*{-1.0cm}    % Move table leftwards
%\protect\footnotesize 
\begin{tabular}{|r|r@{$\,\pm\,$}r|r@{$\,\pm\,$}r|%
                   r@{$\,\pm\,$}r|r@{$\,\pm\,$}r|}% 
\hline\hline  \\[-0.5cm] 
\multicolumn{1}{|c|}{$L$}    & 
\multicolumn{2}{c|}{$\tau_{{\rm int},{\cal M}^2}$} & 
\multicolumn{2}{c|}{$\tau_{{\rm int},{\cal N}}$}   & 
\multicolumn{2}{c|}{$\tau_{{\rm int},{\cal E}}$}   & 
\multicolumn{2}{c|}{$\tau_{{\rm int},{\cal E}^\prime}$} \\[0.1cm] 
\hline\hline 
    4 &     4.013 &   0.018 &     3.205 &   0.013 &     4.023 &   0.018 &     4.034 &   0.018 \\ 
    8 &     5.983 &   0.028 &     5.194 &   0.023 &     6.033 &   0.028 &     6.080 &   0.028 \\ 
   16 &     8.816 &   0.041 &     8.084 &   0.036 &     9.025 &   0.043 &     9.117 &   0.043 \\ 
   32 &    12.648 &   0.057 &    12.200 &   0.055 &    13.280 &   0.062 &    13.438 &   0.063 \\ 
   64 &    18.101 &   0.085 &    18.313 &   0.086 &    19.549 &   0.095 &    19.769 &   0.097 \\ 
  128 &    25.665 &   0.117 &    27.094 &   0.127 &    28.525 &   0.137 &    28.820 &   0.139 \\ 
  256 &    35.691 &   0.164 &    39.204 &   0.189 &    40.824 &   0.200 &    41.212 &   0.203 \\ 
  512 &    49.775 &   0.480 &    56.667 &   0.583 &    58.511 &   0.612 &    58.974 &   0.619 \\ 
 1024 &    68.387 &   1.183 &    80.489 &   1.511 &    82.488 &   1.567 &    83.030 &   1.583 \\ 
[0.1cm] 
\hline\hline
\end{tabular}
\caption{Autocorrelation times for the runs performed at the critical point 
of the 3-state Potts model.
For each lattice size ($L$), we include the integrated autocorrelation times 
for the squared magnetization ($\tau_{{\rm int},{\cal M}^2}$), the bond  
occupation ($\tau_{{\rm int},{\cal N}}$), the energy  
($\tau_{{\rm int},{\cal E}}$), 
and the nearest-neighbor connectivity ($\tau_{{\rm int},{\cal E}^\prime}$).  
} 
\label{table_dynamic_full}
\end{table}

\clearpage  

%% Table generated by sw_potts3_make_table_lisok.pl
%% Data for 3-state Potts model @ Tc
%% TABLE 3: RATIO TAU/CV 
%% 
\begin{table} 
\centering 
\vspace*{-0.5in} % Move table upwards
\addtolength{\tabcolsep}{-1.0mm}
\hspace*{-1.0cm}    % Move table leftwards
%\protect\footnotesize 
\begin{tabular}{|r|rrr|}%
\hline\hline  \\[-0.5cm] 
\multicolumn{1}{|c|}{$L$}    & 
\multicolumn{3}{c|}{$\tau_{{\rm int},{\cal E}}/C_H$}\\[0.1cm] 
\hline\hline 
    4 &   2.6906 & $\pm$ (0.0111, & $\leq$ 0.0161) \\ 
    8 &   2.4413 & $\pm$ (0.0092, & $\leq$ 0.0145) \\ 
   16 &   2.4285 & $\pm$ (0.0090, & $\leq$ 0.0145) \\ 
   32 &   2.5084 & $\pm$ (0.0084, & $\leq$ 0.0147) \\ 
   64 &   2.6476 & $\pm$ (0.0111, & $\leq$ 0.0163) \\ 
  128 &   2.8241 & $\pm$ (0.0111, & $\leq$ 0.0171) \\ 
  256 &   2.9773 & $\pm$ (0.0113, & $\leq$ 0.0184) \\ 
  512 &   3.1670 & $\pm$ (0.0281, & $\leq$ 0.0418) \\ 
 1024 &   3.3630 & $\pm$ (0.0442, & $\leq$ 0.0810) \\ 
[0.1cm] 
\hline\hline
\end{tabular}
\caption{Ratios $\tau_{{\rm int},{\cal E}}/C_H$ for the runs performed at
the critical point of the 3-state Potts model. In parentheses we give our 
error-bar estimates. The first number shows the error bar coming from 
performing the ``bunch'' method with 100 bunches. The second number is 
obtained by using the triangular inequality with the numerical results 
coming from the standard method (see Section~\protect\ref{sec_analysis}).   
}
\label{table_tau_over_cv}
\end{table}

%% Data for 3-state Potts model @ Tc
%% TABLE 4: results 
%%
\begin{table}
\centering
\begin{tabular}{|l|lc|}%
\hline\hline  %\\[-0.5cm]
\multicolumn{1}{|c|}{Exponent} & \multicolumn{1}{|c}{numerical} & exact \\
\hline \hline
$\gamma/\nu$  &$1.73444   \pm 0.00043$ & 26/15 \\ 
$\alpha/\nu$  &$0.4240\kk \pm 0.0030$  & 2/5   \\  
\hline 
$z_{{\rm int},{\cal E}}$
              &$0.515\kk\kk \pm 0.006$  & $\ge 2/5$ \\ 
$z_{{\rm int},{\cal E'}}$
              &$0.514\kk\kk \pm 0.006$  & $\ge 2/5$ \\  
$z_{{\rm int},{\cal N}}$
              &$0.529\kk\kk \pm 0.006$  & $\ge 2/5$ \\  
$z_{{\rm int},{\cal M}^2}$
              &$0.475\kk\kk \pm 0.006$  &    
\\[0.1cm]
\hline\hline
\end{tabular}
\caption{Numerical estimates for the static and dynamic critical exponents 
of the 3-state Potts model (second column). We also include the exact 
results for comparison (third column).  
}
\label{table_results}
\end{table}

%%%%%%%%%%%% START OF FIGURES %%%%%%%%%%%%%
\newpage 

%
% FIGURE 1
%
% autocorrelation function N  
%
%
 
\newpage

%
% FIGURE 2
%
% autocorrelation function for E  
%
%

\newpage 

%
% FIGURE 3
%
% autocorrelation function for E'  
%
%

\clearpage

%%\doublespace
%%\listoffigures

%%\clearpage

%%\doublespace
%%\listoftables

%%\clearpage

\end{document}